\documentclass[prd,twocolumn,showpacs,amsmath,amssymb,superscriptaddress,nofootinbib,showkeys]{revtex4}
\usepackage{amssymb}
\usepackage{amsmath}
\usepackage{graphicx}

\newcommand{\eps}{\varepsilon}
\newcommand{\Fst}{{\mathop {\rule{0pt}{0pt}{F}}\limits^{\;*}}\rule{0pt}{0pt}}
\renewcommand{\Im}{\mathop{\rm Im}\nolimits}
\newcommand{\Real}{\mathop{\rm Re}\nolimits}
\newcommand{\sgn}{\mathop{\rm sgn}\nolimits}
\newcommand{\arccot}{\mathop{\rm arccot}\nolimits}

\begin{document}
\title{Electromagnetic waves in an axion-active relativistic plasma non-minimally coupled to gravity}
\author{Alexander B. Balakin}
\email{Alexander.Balakin@kpfu.ru} \affiliation{Department of
General Relativity and Gravitation, Institute of Physics, Kazan
Federal University, Kremlevskaya str. 18, Kazan 420008, Russia}

\author{Ruslan K. Muharlyamov}
\email{Ruslan.Muharlyamov@kpfu.ru} \affiliation{Department of
General Relativity and Gravitation, Institute of Physics, Kazan
Federal University, Kremlevskaya str. 18, Kazan 420008, Russia}

\author{Alexei E. Zayats}
\email{Alexei.Zayats@kpfu.ru} \affiliation{Department of General
Relativity and Gravitation, Institute of Physics, Kazan Federal
University, Kremlevskaya str. 18, Kazan 420008, Russia}


\begin{abstract}
We consider cosmological applications of a new self-consistent
system of equations, accounting for a nonminimal coupling of the
gravitational, electromagnetic and pseudoscalar (axion) fields in
a relativistic plasma. We focus on dispersion relations for
electromagnetic perturbations in an initially isotropic
ultrarelativistic plasma coupled to the gravitational and axion
fields in the framework of isotropic homogeneous cosmological
model of the de Sitter type. We classify the longitudinal and
transversal electromagnetic modes in an axionically active plasma
and distinguish between waves (damping, instable or running), and
nonharmonic perturbations (damping or instable). We show that for
the special choice of the guiding model parameters the transversal
electromagnetic waves in the axionically active plasma,
nonminimally coupled to gravity, can propagate with the phase
velocity less than speed of light in vacuum, thus displaying a
possibility for a new type of resonant particle-wave interactions.
\end{abstract}



\maketitle

\section{Introduction}

Electromagnetic radiation provides the most impor\-tant channel of
information about our Universe. Valuable information about cosmic
sources of photons, and about cosmic events accompanying the light
propagation, is encoded in the intensity, polarization, phase and
spectral characteristics of the electromagnetic radiation. In this
sense, reconstruction of the phase and group velocities of
observed electromagnetic waves, which travel through plasma and
gas in the dark matter environment, gives us the basis for
theoretical modeling of the properties of these cosmic substrates.
When we deal with plasma, the phase and group velocities are given
by $V_{{\rm ph}}{=}\frac{\omega}{k}$ and $V_{{\rm gr}}{=}
\frac{\partial \omega}{\partial k}$, respectively, where $\omega$
is the frequency and $k$ is the wave three-vector modulus. Thus,
the dependence $\omega=\omega(k)$ obtained from the so-called {\it
dispersion relations} for the longitudinal and transversal
electromagnetic waves in plasma plays an important role in the
information decryption. The theory of dispersion relations is
well-elaborated for various plasma configurations (see, e.g.,
\cite{Silin,2,LLX,4} for details, review and references). To
obtain novel results in this scientific sphere we intend to use a
new nonminimal Einstein-Maxwell-Vlasov-axion model \cite{PlasmaI},
which deals with the self-consistent theory of tidal-type
interactions between gravitational, electromagnetic, pseudoscalar
(axion) fields and a relativistic multi-component plasma.

Cosmological applications of this model seem to be the most
interesting, since the study of the interaction of four key
players of the nonminimal Einstein-Maxwell-Vlasov-axion model
(gravitons, photons, electrically charged plasma particles and
axions) is important for the description of the history of our
Universe.  There are at least three motives, which could explain
this interest. First, in the early Universe the nonminimal couping
of a matter and fields to gravity was important on the stage of
inflation, when the space-time curvature was varying
catastrophically fast. Late-time accelerated expansion discovered
recently \cite{A1,A2,A3} has revived an interest to inertial,
tidal and rip's effects \cite{8,9,10,11}, for which the curvature
coupling could be also important. Second, the cold dark matter,
which now is considered to be one of two key elements of the dark
fluid, guiding the late-time Universe evolution
\cite{DM1,DM2,DM3}, contains (hypothetically) an axion subsystem.
In the early Universe the axions (light pseudo-Goldstone bosons)
could be created due to the phase transition associated with the
Peccei-Quinn symmetry breakdown
\cite{Peccei,8Ni,Weinberg,Wilczek0,Peccei2,Battesti,Raffelt,Turner,Khlopov}.
In the late-time Universe these pseudo-bosons, probably, exist as
relic axions, forming various cold dark matter configurations.
Third, electrically charged plasma and photons are ubiquitous: one
can find plasma configurations and an ocean of photons in every
epoch in the cosmic history and in many objects, which form our
Universe. Of course, studying the nonminimal
Einstein-Maxwell-Vlasov-axion theory, we are restricted to
theoretical modeling of the dispersion relations for relativistic
homogeneous axionically active plasma. For instance, plasma can be
treated as relativistic substratum in the early Universe; however,
as was shown in the papers \cite{K1,K2,K3}, the Peccei-Quinn phase
transition is accompanied by the creation of strongly
inhomogeneous axionic primordial configurations, which were
indicated (see, e.g., \cite{K1,Khlopov}) as ``archioles''. The
inhomogeneity of archioles type is frozen at the radiation
domination stage and should be inherited in the large scale
structure of the modern Universe. Additional problem in the
theoretical modeling is connected with instabilities generated in
plasma due to its inhomogeneity (see, e.g., \cite{Mikh}).
Self-gravitation of the pseudoscalar (axion) field produces
gravitational instability similarly to the instability of scalar
fields described in \cite{K4}, thus making the studying of the
electromagnetic waves propagation much more sophisticated. In such
situation we need of some toy-model, which provides a balance
between complexity of the problem as a whole and mathematical
clarity of the simplified model. For the first step we have chosen
the approximation, which is standard for homogeneous cosmological
model: the space-time metric and the axionic dark matter
distribution are considered to depend on time only. In this case
the standard approach to the analysis of waves propagation in
relativistic plasma is based on the study of dispersion relations
in terms of frequency $\omega$ and wave three-vector $\vec{k}$.
When we consider homogeneous cosmological models, we can use the
standard Fourier transformations with respect to  spatial
coordinates and can naturally introduce the analog of $\vec{k}$.
The Fourier-Laplace transformation with respect to time is,
generally, non-effective, since the coefficients in the master
equations for the electromagnetic field depend on cosmological
time. However, we have found one specific very illustrative model
(the plasma is ultrarelativistic, the spacetime is of a constant
curvature, the axion field has a constant time derivative), for
which the master equations can be effectively transformed into a
set of differential equations with constant coefficients. This
simplified the analysis of the dispersion relations essentially,
and allowed us to interpret the results in the standard terms. As
a next step, we plan to consider homogeneous but anisotropic
models, and then models with inhomogeneous distribution of the
axionic dark matter.

In this work we obtain and analyze the dispersion relations for
perturbations in the axionically active plasma, non-minimally
coupled to gravity, classify these perturbations and distinguish
the transversal waves, which can propagate with the phase velocity
less than the speed of light in vacuum. The work is organized as
follows. In Section~\ref{sec2} we describe the appropriate
background state of the system as a whole, and discuss the exact
solutions of non-perturbed master equations for the plasma in the
gravitational and axionic fields, which were derived in
\cite{PlasmaI}. In Section~\ref{sec3} we consider the equations
for the electromagnetic perturbations in the $(t,\vec{x})$ form
and then the equations for the corresponding Fourier-Laplace
images in the $(\Omega,\vec{k})$ form. We analyze in detail the
dispersion relations for longitudinal waves in Section~\ref{sec4},
and for transversal waves in Section~\ref{sec5}. We summarize the
results in Section~\ref{sec6} (Discussion).

\section{On the background solutions to the nonminimal master
equations}\label{sec2}

Let us consider the {\it background} solutions for the nonminimal
Einstein-Maxwell-Vlasov-axion model, which satisfy the following
four conditions. {\it First}, we assume that the spacetime is
isotropic, spatially homogeneous and is described by the metric
\begin{equation}
ds^2 = a^2(\tau)[d\tau^2-(dx^2+dy^2+dz^2)] \,.
\label{m0}
\end{equation}
{\it Second}, we suppose that the pseudoscalar field inherits the
spacetime symmetry and depends on time only, $\phi(\tau)$. {\it
Third}, we assume, that both the background macroscopic (external)
and cooperative (internal) electromagnetic fields are absent,
i.e., the total Maxwell tensor is equal to zero $F_{ik}=0$. {\it
Fourth}, we consider the plasma to be the test ingredient of the
cosmological model; this means that the contribution of the plasma
particles into the total stress-energy tensor is negligible in
comparison with the dark energy contribution, presented in our
model by the $\Lambda$ term, and with the dark matter one,
described by the axion field $\Psi_0 \phi$ ($\phi$ is
dimensionless pseudoscalar field). These four requirements provide
the master equations obtained in \cite{PlasmaI} to be reduced to
the following system.

\subsection{Background nonminimal equations for the gravity field}

In the absence of the electromagnetic field the nonminimally
extended Einstein equations take the form (see Subsection IIIE in
\cite{PlasmaI})
\begin{gather}
R_{ik}-\frac{1}{2}Rg_{ik} - \Lambda g_{ik} ={}\nonumber \\ {}=
\kappa \Psi^2_0 \left[T^{({\rm A})}_{ik}  + \eta_2 {\cal
T}^{(5)}_{ik} + \eta_3 {\cal T}^{(6)}_{ik} + \eta_{({\rm A})}
{\cal T}^{(7)}_{ik} \right] \,, \label{Eineq}
\end{gather}
where the term
\begin{equation}
T^{({\rm A})}_{ik} = \left\{\nabla_i \phi \nabla_k \phi
-\frac{1}{2} g_{ik} \left[ \nabla^m \phi \nabla_m \phi -
m^2_{({\rm A})} \phi^2 \right] \right\} \label{TAX}
\end{equation}
is the stress-energy tensor of the pseudoscalar field.
The terms
\begin{gather}
{\cal T}^{(5)}_{ik}=  R \nabla_i \phi \nabla_k \phi + \left(g_{ik}
\nabla_n \nabla^n - \nabla_i \nabla_k \right)\left[\nabla_m \phi
\nabla^m \phi \right]+{}\nonumber\\
{}+\nabla_m \phi \nabla^m \phi
\left(R_{ik}-\frac{1}{2}Rg_{ik}\right) \,,\label{calT5}
\end{gather}
\begin{gather}
{\cal T}^{(6)}_{ik} = \nabla_m \phi \left[R_i^m \nabla_k \phi +
R_k^m \nabla_i \phi \right]+{}\nonumber\\
{}+\frac{1}{2}g_{ik}\left( \nabla_m \nabla_n -R_{mn}\right)\left[
\nabla^m \phi \nabla^n \phi \right] - {} \nonumber\\ {}-\nabla^m
\left[ \nabla_m \phi\; \nabla_i \nabla_k \phi \right]
\,.\label{calT6}
\end{gather}
\begin{equation}\label{calT7}
{\cal T}^{(7)}_{ik}= \left(\nabla_i \nabla_k - g_{ik} \nabla_m
\nabla^m \right){\phi}^2 -
\left(R_{ik}-\frac{1}{2}Rg_{ik}\right)\phi^2
\end{equation}
describe the nonminimal contributions associated with coupling
constants $\eta_2$, $\eta_3$, and  $\eta_{({\rm A})}$,
respectively.

We are interested to analyze the specific solution to these
equations, which is characterized by the de Sitter metric
\begin{equation}\label{dS}
ds^2 = \frac{1}{H^2\tau^2}[d\tau^2-(dx^2+dy^2+dz^2)]
\end{equation}
with
\begin{equation}
a(\tau)= -\frac{1}{H\tau}\,, \label{m1}
\end{equation}
where $H$ is a constant. With the transformation of time
\begin{equation}
\tau = -\frac{1}{H a(t_0)} e^{-H(t-t_0)}\,,
\label{m11}
\end{equation}
which gives the correspondence: $\tau \to 0$, when $t \to \infty$;
$\tau \to -\infty$, when $t \to -\infty$;  $\tau \to -\frac{1}{H
a(t_0)} \equiv \tau_0$, when $t \to t_0$, one can obtain the
well-known form of the de Sitter metric
\begin{equation}
ds^2 = dt^2 - e^{2H(t-t_0)}a^2(t_0)(dx^2+dy^2+dz^2) \,.
\label{m01}
\end{equation}
The de Sitter metric (\ref{dS}) describes the spacetime of
constant curvature, for which the basic geometric quantities of
the Riemann tensor, the Ricci tensor and Ricci scalar take the
form
\begin{gather}
R_{ikmn}=-H^2(g_{im}g_{kn}-g_{in}g_{km})\,,\nonumber\\
R_{im}=-3H^2g_{im}\,,\quad R=-12H^2\,, \label{m2}
\end{gather}
and the nonminimal susceptibility tensors
\begin{align}
{\cal R}^{ikmn} &\equiv  \frac{q_1}{2} R
(g^{im}g^{kn}-g^{in}g^{km})
+{}\nonumber\\
{}&+\frac{q_2}{2} (R^{im}g^{kn} {-} R^{in}g^{km} {+} R^{kn}g^{im}
{-} R^{km}g^{in})+{}\nonumber\\
{}&+ q_3 R^{ikmn}\,,
\end{align}
\begin{align}
{\chi}^{ikmn}_{(A)} &\equiv  \frac{Q_1}{2} R
(g^{im}g^{kn}-g^{in}g^{km})
+{}\nonumber\\
{}&+\frac{Q_2}{2} (R^{im}g^{kn} {-} R^{in}g^{km} {+} R^{kn}g^{im}
{-} R^{km}g^{in})+{}\nonumber\\
{}&+ Q_3 R^{ikmn}\,,
\end{align}
\begin{gather}
\Re^{mn}_{({\rm A})} \equiv \frac{1}{2} \eta_1
\left(F^{ml}R^{n}_{\ l} + F^{nl}R^{m}_{\ l} \right) + \eta_2 R
g^{mn} + \eta_3 R^{mn}
\end{gather}
introduced in \cite{PlasmaI} transform into
\begin{equation}
{\cal R}^{ikmn}=-H^2(6q_1+3q_2+q_3) \ (g^{im}g^{kn}-g^{in}g^{km})\,,
\label{m3}
\end{equation}
\begin{equation}
{\chi}^{ikmn}_{\rm (A)}=-H^2(6Q_1+3Q_2+Q_3) \
(g^{im}g^{kn}-g^{in}g^{km})\,, \label{m4}
\end{equation}
\begin{equation}
\Re^{mn}_{(A)}=-3H^2(4\eta_2+\eta_3)g^{mn} \,.
\label{m5}
\end{equation}
One can check directly that the de Sitter-type metric (\ref{m0})
with the scale factor (\ref{m1}) is the exact solution of the
equations (\ref{Eineq})-(\ref{calT7}) with the pseudoscalar field
linear in time, i.e., $\phi(\tau){=}\nu \tau$, when the following
three relationships are satisfied:
\begin{equation}
\Lambda = 3H^2 \,, \quad m^2_{({\rm A})} = \frac{2}{9} \Lambda^2(6\eta_2+\eta_3) \,,
\label{m6}
\end{equation}
\begin{equation}
\eta_{({\rm A})} = - \frac{1}{6} + \frac{1}{9} \Lambda (9\eta_2+2\eta_3) \,.
\label{m7}
\end{equation}
These relations contain neither $\nu$, nor $\Psi_0$. In addition,
when $\eta_2=\eta_3=0$, we obtain the result
\begin{equation}
\Lambda = 3H^2 \,, \quad m_{({\rm A})} = 0 \,, \quad \eta_{({\rm
A})} = - \frac{1}{6} \,, \label{m77}
\end{equation}
which is well-known for the massless scalar field conformally
coupled to gravity \cite{Penrose}. When the constants of
nonminimal coupling $\eta_2$ and $\eta_3$ are nonvanishing, we can
consider the axions to be massive, and the mass $m_{({\rm A})}$
itself is connected not only with these coupling constants, but
with the cosmological constant $\Lambda$ as well. In order to
minimize the number of unknown coupling parameters, one can put,
for instance,
\begin{equation}
\eta_3 = - \frac{9}{2}\eta_2\,,  \quad \eta_{({\rm A})} = - \frac{1}{6} \,, \quad \eta_2 = \frac{3 m^2_{({\rm A})}}{\Lambda^2} \,,
\label{m78}
\end{equation}
satisfying (\ref{m6}) and (\ref{m7}).

\subsection{Nonminimal equations for the background pseudoscalar (axion) field}

In the absence of the background electromagnetic field the
nonminimally extended master equation (see Eq.(88) in
\cite{PlasmaI}) for the pseudoscalar $\phi$ takes the form
\begin{gather}
\nabla_m \left[ \left( g^{mn} + \Re^{mn}_{({\rm A})} \right)
\nabla_n \phi \right] {+} \left[m^2_{({\rm A })} {+} \eta_{({\rm
A})} R \right] \phi ={}\nonumber\\ {}=- \frac{1}{\Psi^2_0}
\sum_{({\rm a})} \int dP f_{({\rm a})} {\cal G}_{({\rm a})} \,.
\label{eqaxi1}
\end{gather}
Let us take into account the relations (\ref{m3})-(\ref{m5}) for
$\Re^{mn}_{({\rm A})}$ and ${\chi}^{ikmn}_{({\rm A})}$, and
suppose that for the background distribution function its
zero-order moment in the right-side hand of (\ref{eqaxi1})
vanishes, then we obtain the equation
\begin{equation}
    [1-3H^2(4\eta_2+\eta_3)]\nabla^m\nabla_m\phi+(m^2_A-12H^2\eta_{({\rm A})})\phi=0 \,.
    \label{eqaxi11}
\end{equation}
The function $\phi=\nu \tau$ satisfies (\ref{eqaxi11}), if the
following relation is valid
\begin{equation}
-2H^2[1-3H^2(4\eta_2+\eta_3)]+m^2_A-12H^2\eta_{({\rm A})}=0 \,.
    \label{eqaxi12}
\end{equation}
We used here the relation
\begin{equation}
\nabla^m\nabla_m\phi=
\frac{1}{a^4}\,\frac{d}{d\tau}\left(a^2\dot\phi\right)=-2H^2\nu \tau  \,.
    \label{eqaxi13}
\end{equation}
Clearly, the relation (\ref{eqaxi12}) is identically satisfied,
when the equations (\ref{m6}), (\ref{m7}) are valid. In other
words, we have shown that the master equation for the background
pseudoscalar (axion) field admits the existence of the exact
solution $\phi(\tau)=\nu \tau$ linear in time, when the spacetime
is of the de Sitter type with the metric (\ref{dS}).

\subsection{Consistency of the background electrodynamic equations}

We assume that the Maxwell tensor describing the background
cooperative electromagnetic field in plasma is equal to zero,
$F_{ik}=0$. It is possible, when the tensor of spontaneous
polarization-magnetization ${\cal H}^{ik}$ vanishes, and when  the
four-vector of electric current in plasma, $I^i$, is equal to
zero. Since $\phi=\phi(\tau)$ and $R^{\alpha 0}=0$ ($\alpha
=1,2,3$) the formula (see (87) in \cite{PlasmaI})
\begin{equation}
{\cal H}^{ik} \equiv -\frac{1}{2}\eta_1 \Psi^2_0 \left[
\left(R^{km} \nabla^i \phi - R^{im} \nabla^k \phi \right) \nabla_m
\phi \right] \,, \label{current}
\end{equation}
gives ${\cal H}^{ik}=0$. The condition $I^i=0$ provides the
restrictions for the background distribution function $f_{({\rm
a})}$, which we will discuss in the next subsection.

\subsection{Background solution to the Vlasov equation}

In the context of the Vlasov theory we search for a 8-dimensional
one-particle distribution function $f_{({\rm a})}(x^i, p_k)$ which
describes particles of a sort $({\rm a})$ with the rest mass
$m_{({\rm a})}$, electric charge $e_{({\rm a})}$. This function of
the coordinates $x^i$ and of the momentum four-covector $p_k$
satisfies the relativistic kinetic equation, which can be
presented now in the form
\begin{equation}
p^i\left(\frac{\partial}{\partial x^i}+\Gamma_{ik}^l p_l\frac{\partial}{\partial p_k}\right)f_{({\rm a})}=0 \,.
\label{kin1}
\end{equation}
We assume here that the background macroscopic cooperative
electromagnetic field in plasma is absent, and there are no
contact interactions between axions and plasma particles, so that
the force ${\cal F}^i_{({\rm a})}$ (see Eq.(29) in \cite{PlasmaI})
disappears from this equation. In the cosmological context we
consider the distribution function depending on the covariant
components of the particle four-momentum, $p_k$ in order to
simplify the analysis. Here $\Gamma^m_{ik}$ are the Christoffel
symbols associated with the spacetime metric $g_{ik}$. The
characteristic equations associated with the kinetic equation
(\ref{kin1})
\begin{equation}
\frac{dp_k}{ds}- \Gamma_{ik}^l p_l \frac{dx^i}{ds} =0 \,, \quad \frac{dx^i}{ds} = \frac{1}{m_{({\rm a})}} g^{ij} p_j  \,,
\label{kin2}
\end{equation}
are known to be directly solved for the metric (\ref{m0}). For the
cases $i=\alpha=1,2,3$ one obtains immediately that
\begin{equation}
p_{\alpha} = q_{\alpha}\,,
\label{kin3}
\end{equation}
where $q_{\alpha}{=} {\rm const}$ are the integrals of motion. The
component $p_0$ of the particle momentum four-vector can be found
from the quadratic integral of motion $g^{ij}p_ip_j=m^2_{({\rm
a})}{=}{\rm const}$:
\begin{equation}
p_{0} = \sqrt{m^2_{({\rm a})} a^2(\tau) + q^2} , \quad q^2 \equiv
q^2_{1} + q^2_{2} + q^2_{3} = {\rm const}. \label{kin4}
\end{equation}
Other three integrals of motion are
\begin{equation}
X^{\alpha } = x^{\alpha}(\tau) + \frac{q_{\alpha}}{q^2}
\sqrt{\frac{m^2_{({\rm a})}}{H^2}+ q^2 \tau^2} \,. \label{kin5}
\end{equation}
Generally, the distribution function, which satisfies kinetic
equation (\ref{kin1}) can be reconstructed as an arbitrary
function of seven integrals of motion $q_{\alpha}$, $X^{\alpha}$,
$\sqrt{g^{ij}p_ip_j}$, nevertheless, taking into account that the
spacetime is isotropic and homogeneous, we require that the
distribution function inherits this symmetry and thus it has the
form
\begin{equation}
f_{({\rm a})}(x^l, p_k) = f^{(0)}_{({\rm a})}(q^2) \delta(\sqrt{g^{sn}p_sp_n}-m_{({\rm a})}) \,,
\label{kin6}
\end{equation}
where $f^{(0)}_{({\rm a})}(q^2)$ is arbitrary function of one argument, namely, $q^2$.
We are interested now to calculate the first moment of the distribution function
\begin{gather}
N_{({\rm a})}^i \equiv \frac{1}{m_{({\rm a})}}\int
\frac{d_4P}{\sqrt{-g}}g^{ij}p_j f_{({\rm a})}(x^l, p_k)
\times{}\nonumber\\
{}\times\delta(\sqrt{g^{sn}p_s p_n} {-}m_{({\rm a})})\Theta(V^h
p_h) \,, \label{kin7}
\end{gather}
where $d_4P=dp_0dp_1dp_2dp_3$ symbolizes the volume in the
four-dimensional momentum space; the delta function guarantees the
normalization property of the particle momentum, the Heaviside
function $\Theta(V^h p_h)$ rejects negative values of energy,
$V^h$ is the velocity four-vector of the system as a whole. In the
de Sitter spacetime with $V^h{=}\delta^h_0 \frac{1}{a(\tau)}$ the
quantities $N_{({\rm a})}^i$ can be reduced to
\begin{equation}
N^i_{({\rm a})}(\tau) =  \int \frac{dq_1 dq_2 dq_3}{a^2 p_0}g^{ij}p_j f^{(0)}_{({\rm a})}(q^2) \,,
\label{kin8}
\end{equation}
therefore, the spatial components $N_{({\rm a})}^{\alpha}$ vanish. The component $N_{({\rm a})}^0$ reads
\begin{equation}
N_{({\rm a})}^0(\tau) =  \frac{4 \pi}{a^4} \int\limits_0^{\infty}
q^2 dq f^{(0)}_{({\rm a})}(q^2) \equiv \frac{{\cal N_{({\rm
a})}}}{a^4(\tau)} \,, \label{kin9}
\end{equation}
where ${\cal N_{({\rm a})}}$ does not depend on time. This means that the four-vector of the electric current in the electro-neutral plasma
\begin{equation}
I^i(\tau) = \sum_{({\rm a})} e_{({\rm a})} N_{({\rm a})}^i = \frac{1}{a^4(\tau)} \delta^i_0 \sum_{({\rm a})} e_{({\rm a})} {\cal N_{({\rm a})}} = 0\,,
\label{kin10}
\end{equation}
is equal to zero at arbitrary time moment $\tau$, if it was equal to zero at the initial time moment. This guarantees that the Maxwell equations are self-consistent.

\section{Electromagnetic perturbations in an axionically active ultrarelativistic plasma nonminimally coupled to
gravity}\label{sec3}

\subsection{Evolutionary equations}

Let us consider now the state of plasma perturbed by a local
variation of electric charge. As usual, we assume, first, that the
distribution function takes the form $f_{({\rm a})} \to
f^{(0)}_{({\rm a})}(q^2) +  \delta f_{({\rm a})}(\tau, x^{\alpha},
p_k)$, second, that the tensor $F_{ik}$ describing the variation
of the cooperative electromagnetic field in plasma is not equal to
zero. Let us stress that the electromagnetic source in the
right-hand side of the master equation for the pseudoscalar field
(see Eq.(88) in \cite{PlasmaI}) is quadratic in the Maxwell
tensor, and thus, in the linear approximation the background axion
field $\phi(\tau)=\nu \tau$ remains unperturbed. Similarly, the
nonminimally extended equations for the gravity field (see
Eqs.(89)-(100) in \cite{PlasmaI}) are considered to be
unperturbed. This is possible, when the term $\eta_1 {\cal
T}^{(4)}_{ik}$, which is in fact the exclusive term linear in the
Maxwell tensor, is vanishing. Below we assume, that $\eta_1=0$,
thus guaranteeing that the background de Sitter spacetime is not
perturbed in the linear approximation. For the perturbed
quantities $F_{ik}$ and $\delta f_{({\rm a})}$ we obtain the
following coupled system of equations:
\begin{gather}
[1-2H^2(6q_1+3q_2+q_3)]\nabla_k
F^{ik}+{}\nonumber\\{}+[1-4H^2(3Q_1-Q_3)] \Fst^{ik} \nabla_k \phi
={}\nonumber\\ {}=-4\pi \sum_{({\rm a})}  e_{({\rm a})} \int dP\,
\delta f_{({\rm a})} g^{ik} p_k \,,\\
\nabla_k \Fst^{ik} =0 \,, \label{eld1}\\
g^{ij}p_j\left(\frac{\partial}{\partial x^i}+\Gamma_{ik}^m
p_m\frac{\partial}{\partial p_k} \right)\delta f_{({\rm a})} =
e_{({\rm a})} p_i F^i_{\ k}\frac{\partial}{\partial p_k}
f^{(0)}_{({\rm a})} \,. \label{eld2}
\end{gather}
Below we analyze this system for the case, when the plasma
particles are ultrarelativistic. The procedure of derivation of
the dispersion relations for the case of the ultrarelativistic
plasma is very illustrative in our model. Below we assume that in
average $q^2 \gg m^2_{({\rm a})} a^2(\tau)$, so that $p_{0}$ can
be replaced by $p_{0} \to q$ in the integrals. Keeping in mind
that in the linear approximation $\phi =\nu \tau =\frac{\nu}{H
a(\tau)}$, one can check directly that the scale factor $a(\tau)$
disappears from the reduced electrodynamic equations, which take
the following form:
\begin{gather}
(1+2K_1)\,\eta^{\alpha\beta}\partial_\alpha F_{\beta 0}=4\pi
\sum_{({\rm a})} e_{({\rm a})} \int d_3q\, \delta f_{({\rm a})}
\,, \label{eld3}\\
(1+2K_1)\,\left[\partial_0F_{\alpha
0}+\eta^{\beta\gamma}\partial_\beta F_{\alpha
\gamma}\right]-{}\nonumber\\ {}-\frac12 \nu
{\epsilon_\alpha}^{\beta\gamma}F_{\beta\gamma}(1+2K_2)={}\nonumber\\
{}= -4\pi \sum_{({\rm a})} e_{({\rm a})} \int \frac{ d_3q \
q_\alpha}{q}\, \delta f_{({\rm a})} \,, \label{eld4}\\
\partial_0 F_{\alpha\beta}+\partial_\alpha F_{\beta
0}- \partial_\beta F_{\alpha 0}=0 \,, \label{eld5}\\
\partial_\alpha
F_{\beta\gamma}+\partial_\beta F_{\gamma \alpha}+ \partial_\gamma
F_{\alpha \beta}=0 \,.
\label{eld6}
\end{gather}
Here we used the following notations:
\begin{align}
K_1 &\equiv -H^2(6q_1{+}3q_2{+}q_3),\nonumber \\
K_2 &\equiv
-2H^2(3Q_1{-}Q_3) \,, \label{eld7}
\end{align}
and introduced the three-dimensional Levi-Civita symbol
\begin{equation}
\epsilon^{\alpha\beta\gamma} \equiv
E^{0\alpha\beta\gamma} \,,
\label{eld8}
\end{equation}
with $\epsilon^{123}=1$. Here and below for the operation with the
indices we use the Minkowski tensor $\eta_{ik}{=} {\rm diag}
(1,{-}1, {-}1, {-}1)$.

The kinetic equation (\ref{eld2}) for the ultrarelativistic plasma
in the linear approximation can be written in the form
\begin{equation}
q \partial_0 \, \delta f_{({\rm a})} +\eta^{\alpha \beta} q_\alpha
\,\partial_\beta\, \delta f_{({\rm a})} = e_{({\rm a})}
\eta^{\alpha \beta} F_{\beta 0} \ q_\alpha \ \frac{df^{(0)}_{({\rm
a})}}{dq} \,. \label{eld9}
\end{equation}
Surprisingly, in the given representation all the electrodynamic
equations (\ref{eld3})-(\ref{eld6}) and the kinetic equation
(\ref{eld9}) look like the set of integro-differential equations
with the coefficients, which do not depend on time. This fact
allows us to apply the method of Fourier transformations, which is
widely used in case, when we deal with the Minkowski spacetime
\cite{Silin,2,LLX,4}.

\subsection{Equations for Fourier-Laplace images}

In order to study in detail perturbations arising in plasma we
consider the Fourier-Laplace transformations (the Fourier
transformation with respect to the spatial coordinates $x^\alpha$
and the Laplace transformation for $\tau$) in the following form:
\begin{equation}
\delta f_{({\rm a})}=\int \frac{d\Omega\, d_3k}{(2\pi)^4}\, \delta\varphi_{({\rm a})}(\Omega,k_\gamma,q_\beta)\, {\rm e}^{i(k_\alpha x^\alpha -\, \Omega \tau)} \,,
\label{eld10}
\end{equation}
\begin{equation}
F_{lk}=\int \frac{d\Omega\, d_3k}{(2\pi)^4}\, {\cal F}_{lk}(\Omega,k_\gamma)\, {\rm e}^{i(k_\alpha x^\alpha -\, \Omega \tau)} \,.
\label{eld11}
\end{equation}
As usual, we assume that $k_1, k_2, k_3$ are pure real quantities
in order to guarantee that both exponential functions, ${\rm e}^{i
k_\alpha x^\alpha}$ and ${\rm e}^{-i k_\alpha x^\alpha}$,  are
finite everywhere. The quantity $\Omega$ is in general case the
complex one, $\Omega{=}\omega{+}i\gamma$, and, as usual, we assume
that perturbations are absent at $\tau<\tau_0$, where $\tau_0
=-\frac{1}{H a(t_0)}$ relates to the moment $t=t_0$ according to
(\ref{m11}). Our choice of the sign minus in the expression for
the phase $\Theta \equiv k_\alpha x^\alpha -\, \Omega \tau$ in the
exponentials in (\ref{eld10}), (\ref{eld11}) can be motivated as
follows. When the value $t-t_0$ is small, the cosmological time
$\tau$ (see (\ref{m11})) can be estimated as $\tau \to -
\frac{1}{H a(t_0)} [1 -H(t-t_0)]$, and thus the phase $\Theta$
reads
\begin{equation}
\Theta = \Theta_0 + k_\alpha x^\alpha -  \frac{\Omega}{a(t_0)} (t-t_0)  \,, \quad  \Theta_0 = -\frac{\Omega}{H a(t_0)}
\,.
\label{m119}
\end{equation}
Using the notation $k_0 =-\frac{\Omega}{a(t_0)}$ we obtain from
(\ref{m119}) the expression for the phase $\Theta =\Theta_0 + k_m
x^m $, which is standard for the case of Minkowski spacetime.
Finally, the term ${\rm e}^{i \Theta}$ has the multiplier ${\rm
e}^{-i \Omega \tau}$ which contains ${\rm e}^{\gamma \tau} \to
{\rm e}^{\frac{\gamma}{a(t_0)} (t-t_0)}$. In other words, both in
terms of $t$ and $\tau$ the quantity $\gamma$ has the same sense:
when $\gamma<0$ we deal with the plasma-wave damping and
$|\gamma|$ is the decrement of damping; when $\gamma>0$ we deal
with increasing of the perturbation in plasma and this  positive
$\gamma$ is the increment of instability.

We should introduce an initial value of the perturbed distribution
function at the moment $\tau{=}\tau_0$, indicated as $\delta
f_{({{\rm a}})}(0,k_\alpha,q_\gamma)$; as for initial data for the
electromagnetic field, we can put without loss of generality that
$F_{\alpha 0}(\tau{=}0,k_{\beta}){=}0$. Using
(\ref{eld3})-(\ref{eld6}) and (\ref{eld9}) the equations for the
Fourier images $\delta\varphi_{({\rm
a})}(\Omega,k_\gamma,q_\beta)$ and ${\cal
F}_{lk}(\Omega,k_\gamma)$ can be written as follows:
\begin{gather}
\left(\Omega-\frac{k_\alpha
q^\alpha}{q}\right)\delta\varphi_{({\rm
a})}={}\nonumber\\{}=i\,e_{({\rm a})}\,{\cal F}_{\alpha
0}\,\frac{q^\alpha}{q}\cdot\frac{d f^{(0)}_{({\rm a})}}{d q} + i\,
\delta f_{({{\rm a}})}(0, k_\alpha,q_\gamma)\,, \label{I}\\
(1+2K_1)\,k^\alpha\,{\cal F}_{\alpha 0}=-4\pi i\sum_{({\rm a})}
e_{({\rm a})} \int d_3q\, \delta\varphi_{({\rm a})} \,,
\label{II}\\
(1+2K_1)\,\left(\Omega\, {\cal F}_{\alpha 0}-k^\gamma\, {\cal
F}_{\alpha \gamma}\right)-\frac{i}{2}\; \nu
{\epsilon_\alpha}^{\beta\gamma}{\cal
F}_{\beta\gamma}(1+2K_2)={}\nonumber\\{}= -4\pi i\sum_{({\rm a})}
e_{({\rm a})} \int d_3q\, \delta\varphi_{({\rm a})}
\frac{q_\alpha}{q}\,, \label{III}\\
{\cal F}_{\alpha\beta}= \Omega^{-1} \left(k_\alpha\,{\cal
F}_{\beta 0}-k_\beta\,{\cal F}_{\alpha 0}\right) \,, \label{IV}\\
k_\alpha\,{\cal F}_{\beta \gamma}+k_\beta\,{\cal F}_{\gamma\alpha}+k_\gamma\,{\cal F}_{\alpha \beta}= 0\,.\label{VI}
\end{gather}
Here and below we use the notation $q^{\alpha} \equiv
\eta^{\alpha\beta}q_{\beta}$ for the sake of simplicity.  Clearly,
the equation (\ref{VI}) is satisfied identically, if we put ${\cal
F}_{\alpha\beta}$ from (\ref{IV}). Then we use the standard
method: we take $\delta\varphi_{({\rm a})}$ from (\ref{I}), put it
into (\ref{II}) and (\ref{III}), use (\ref{IV}) and obtain,
finally, the equations for the Fourier images of the components
${\cal F}_{\beta 0}$ of the Maxwell tensor:
\begin{gather}
\left[k^\alpha (1+2K_1)-{}\vphantom{\frac{df^{(0)}_{({\rm
a})}}{dq}}\right.\nonumber\\
\left.{}-4\pi \sum_{({\rm a})} e_{({\rm a})}^2 \int \frac{d_3q \,
q^\alpha}{(q\Omega-k_\beta q^\beta)}\cdot\frac{df^{(0)}_{({\rm
a})}}{dq}\right]{\cal F}_{\alpha 0}={\cal J}_0\,,\label{Ipl}
\end{gather}
\begin{gather}
\left\{(1+2K_1)\left[\Omega^2 \delta_\alpha^\gamma -
k^2\,\Pi_\alpha^\gamma\right]- i\;\nu
{\epsilon_\alpha}^{\beta\gamma}k_\beta (1{+}2K_2) -
{}\vphantom{\frac{df^{(0)}_{({\rm a})}}{dq}}\right.\nonumber\\
\left.{}-4\pi \sum_{({\rm a})} \Omega e_{({\rm a})}^2 \int
\frac{d_3q \, q_\alpha q^\gamma}{q (q\Omega-k_\beta
q^\beta)}\cdot\frac{df^{(0)}_{({\rm a})}}{dq}\right\}{\cal
F}_{\gamma 0}= \Omega {\cal J}_\alpha \,.\label{IIpl}
\end{gather}
The Fourier images of the initial perturbations of the electric
current ${\cal J}_0$ and ${\cal J}_\alpha$ are defined as follows:
\begin{equation}
{\cal J}_0 \equiv
4\pi \sum_{({\rm a})} e_{({\rm a})} \int \frac{d_3q \, q \ \delta f_{({\rm a})}(0,k_\beta,q_\gamma)}{q\Omega - k_\mu q^\mu}
\,,\label{XI}
\end{equation}
\begin{equation}
{\cal J}_\alpha \equiv
4\pi \sum_{({\rm a})} e_{({\rm a})} \int \frac{d_3q \, q_\alpha \ \delta f_{({\rm a})}(0,k_\beta,q_\gamma)}{q\Omega - k_\mu q^\mu}
\,.\label{XII}
\end{equation}
The compatibility condition for the current four-vector $\nabla_k {\cal J}^k {=}0$, written in terms of Fourier images
\begin{equation}
\Omega {\cal J}_0 - k^\alpha {\cal J}_\alpha = 4\pi \sum_{({\rm
a})} e_{({\rm a})} \int d_3q  \,\delta f_{({\rm
a})}(0,k_\beta,q_\gamma) = 0 \,,\label{XIII}
\end{equation}
requires in fact that the perturbation in the plasma state does not change the particle number.
The term $\Pi_\alpha^\gamma$ is a projector:
\begin{equation}
\Pi_\alpha^\gamma = \delta_\alpha^\gamma+\frac{k_\alpha k^\gamma}{k^2}
\,, \quad  \Pi_\alpha^\gamma \Pi_\gamma^\mu = \Pi_\alpha^\mu
\,, \label{Ipl6}
\end{equation}
it is orthogonal to $k_{\alpha}$, i.e.,
\begin{equation}
\Pi_\alpha^\gamma k^{\alpha}= 0 = \Pi_\alpha^\gamma k_{\gamma}
\,.\label{Ip46}
\end{equation}
The quantity $k^2$ is defined as $k^2=-k_\beta k^\beta$; it is real and positive.

\subsection{Permittivity tensors}

As usual, we introduce the standard permittivity tensor for the
spatially isotropic relativistic plasma \cite{Silin}
\begin{equation}
\eps_\alpha^\gamma = \delta_\alpha^\gamma - \frac{4\pi }{\Omega}\sum_{({\rm a})}
e_{({\rm a})}^2 \int \frac{d_3q\, q_\alpha q^\gamma}{q(q\Omega-k_\beta q^\beta)}\cdot\frac{df^{(0)}_{({\rm a})}}{dq}\,,
\label{e1}
\end{equation}
and obtain the decomposition
\begin{equation}
\eps_\alpha^\gamma=\eps_\bot\left(\delta_\alpha^\gamma+\frac{k_\alpha
k^\gamma}{k^2}\right)-\eps_{||}\cdot\frac{k_\alpha  k^\gamma}{k^2}\,,
\label{e2}
\end{equation}
where $\eps_\bot$ and $\eps_{||}$ are the scalar transversal and
longitudinal permittivities, respectively:
\begin{equation}
\eps_\bot \equiv \frac{1}{2}\left(\eps^\alpha_\alpha - \eps_{||} \right)
\,, \label{e3}
\end{equation}
\begin{equation}
\eps_{||} \equiv 1 + \frac{4\pi}{k^2}\sum_{({\rm a})}
e_{({\rm a})}^2 \int \frac{d_3q\, k_\alpha q^\alpha}{(q\Omega-k_\beta q^\beta)}\cdot\frac{df^{(0)}_{({\rm a})}}{dq}\,.
\label{e4}
\end{equation}
Finally, we decompose the Fourier image of the electric field
${\cal F}_{\gamma 0}$ into the longitudinal and transversal
components with respect to the wave three-vector
\begin{equation}
{\cal F}_{\gamma 0} = \frac{k_\gamma}{k} \ {\cal E}^{||} + {\cal E}^{\bot}_\gamma \,,
\label{Edecomp1}
\end{equation}
where
\begin{equation}
{\cal E}^{||} \equiv - {\cal F}_{\gamma 0} \frac{k^{\gamma}}{k} \,, \quad {\cal E}^{\bot}_\gamma \equiv {\cal F}_{\beta 0} \Pi^{\beta}_{\gamma} \,,
\label{Edecomp2}
\end{equation}
and obtain the split equations for the Fourier images of the
longitudinal and transversal electric field components,
respectively:
\begin{equation}
{\cal E}^{||}= - \frac{{\cal J}_0}{k \left(\eps_{||}{+}2K_1
\right)}  \,, \label{Ipq}
\end{equation}
\begin{gather}
\left[\left(\eps_\bot+2K_1-\frac{(1{+}2K_1)k^2}{\Omega^2}\right)\delta_\alpha^\gamma
+ {}\vphantom{\frac{i\; \nu}{\Omega^2}}\right.\nonumber\\
\left.{}+\frac{i\; \nu}{\Omega^2}\;{\epsilon_\alpha}^{\gamma
\beta}k_\beta (1{+}2K_2)\right] {\cal E}_{\gamma}^\bot=
\frac{1}{\Omega} \Pi^\gamma_\alpha {\cal J}_\gamma \,.
\label{IIpq}
\end{gather}
The second term in the brackets (linear in the wave three-vector
$k_\beta$) describes the well-known effect of optical activity:
two transversal components of the field ${\cal E}_{\gamma}^\bot$
are coupled, so that linearly polarized waves do not exist, when
$\nu (1{+}2K_2)\neq 0$. Since the so-called gyration tensor
$\frac{\nu}{\Omega^2}\,{\epsilon_\alpha}^{\gamma \beta}
(1{+}2K_2)$ is proportional to the Levi-Civita symbol
${\epsilon_\alpha}^{\gamma \beta}$, we deal with {\it natural
optical activity} according to the terminology used in
\cite{LLVIII}, which can be described by one (pseudo) scalar
quantity. Thus, we can speak about axionically induced optical
activity in plasma, and about axionically active plasma itself.

\subsection{Dispersion relations}

The inverse Fourier-Laplace transformation (\ref{eld11}) of the
electromagnetic field is associated with the calculation of the
residues in the singular points of two principal types. First, one
should analyze the poles of the functions ${\cal J}_0$ and ${\cal
J}_\alpha$ (see (\ref{XI}) and \ref{XII})) describing initial perturbations;
the most known among
them are the Van Kampen poles $\Omega{=}\frac{k_\alpha
q^\alpha}{q}$. The poles of the second type appear as the roots of
the equations
\begin{equation}
\eps_{||}+2K_1=0 \,,
\label{longDR}
\end{equation}
and
\begin{gather}
\det\left[\left(\eps_\bot+2K_1-\frac{(1{+}2K_1)k^2}{\Omega^2}\right)\delta_\alpha^\gamma
- {}\vphantom{\frac{i\;
\nu}{\Omega^2}}\right.\nonumber\\\left.{}-\frac{i\;
\nu}{\Omega^2}\;{\epsilon_\alpha}^{\beta\gamma}k_\beta
(1{+}2K_2)\right] = 0 \,. \label{IX}
\end{gather}
The equation (\ref{longDR}) is the nonminimal generalization of
the dispersion relations for the longitudinal plasma waves; it
includes the spacetime curvature and the constants of nonminimal
coupling (via the term $K_1$, see (\ref{eld7})), nevertheless, it
does not contain any information about the axion field. The
equation (\ref{IX}) describes transversal electromagnetic waves in
axionically active plasma nonminimally coupled to gravity; it can
be transformed into
\begin{gather}
\left[\eps_\bot{+}2K_1{-}\frac{(1{+}2K_1)k^2}{\Omega^2}\right]
\left[\left(\eps_\bot {+}2K_1 {-}
\frac{(1{+}2K_1)k^2}{\Omega^2}\right)^2 -{}\right.\nonumber\\
\left.{}-\frac{\nu^2(1{+}2K_2)^2k^2}{\Omega^4}\right]= 0.
\label{trans1}
\end{gather}
This dispersion equation generalizes the one obtained in \cite{Itin1,Itin2} for a minimal axionic vacuum. Clearly, one should consider two important cases.

\subsubsection{Special case $1+2K_2=0$}

The condition $1{+}2K_2=0$ rewritten as $3Q_1{-}Q_3 {=}\frac{1}{4H^2}$, provides the dispersion relations to be of the form
\begin{equation}
\left[\eps_\bot{+}2K_1{-}\frac{(1{+}2K_1)k^2}{\Omega^2}\right]{=} 0 \,,
\label{trans11}
\end{equation}
includes the curvature terms and does not contain the information about the axion field.

\subsubsection{General case $1+2K_2 \neq 0$}

In this case the dispersion relations for the transversal electromagnetic perturbations read
\begin{equation}
\eps_\bot = - 2K_1+\frac{(1+2K_1)k^2\pm \nu (1+2K_2)k}{\Omega^2} \,,
\label{transDR}
\end{equation}
displaying explicitly the dependence on the axion field strength
$\nu$. Two signs, plus and minus, symbolize the difference in the
dispersion relations for waves with left-hand and right-hand polarization
rotation. In this sense, when $1{+}2K_2 \neq 0$, we deal with an
axionically active plasma.

\subsection{Analytical properties of the permittity tensor  and the inverse Laplace transformation}

\begin{figure}[h]
  \includegraphics[width=7cm]{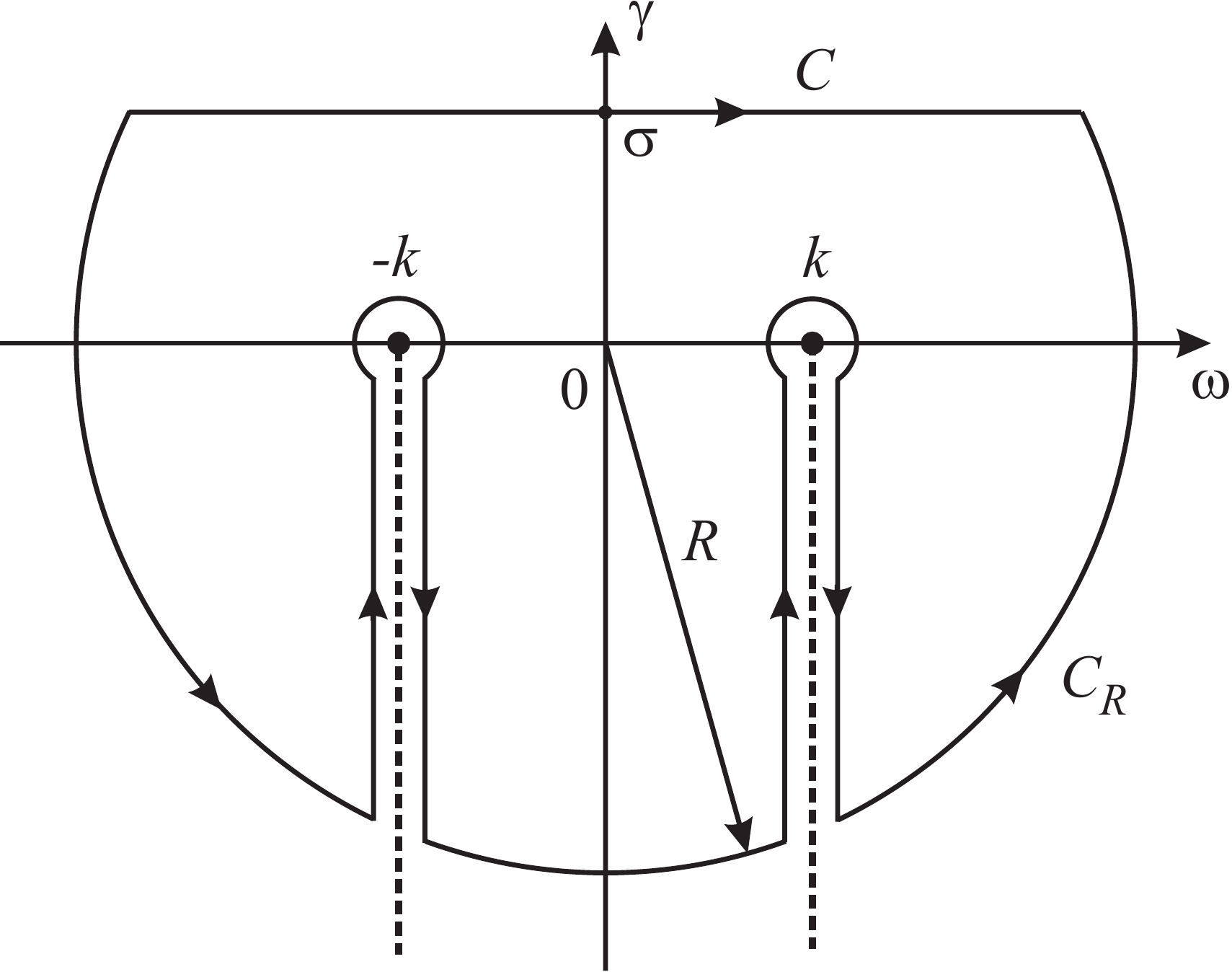}\\
  \caption{Integration contour $C_{{\rm R}}$ for the inverse Laplace
transformation of the electromagnetic field strength.  The radius
$R{=}\sqrt{A^2{+}\sigma^2}$ tends to infinity, providing that all
the poles of the function ${\cal F}_{lk}(\Omega,k_\gamma)$ are
inside. The branchpoints $\Omega{=}\pm k$ are excluded by using
cuts along the lines $\omega {=} \pm k$, $\gamma {=} k y$,
$-\infty<y <0$, and infinitely small circles harboring the
branchpoints. } \label{Contur}
\end{figure}

Let us remind three important features concerning the Laplace
transformation in the context of relativistic plasma theory.
First, as usual, we treat this transformation as a limiting
procedure
\begin{equation}
f(\tau)=\frac{1}{2\pi}\lim_{A\to+\infty}\int\limits_{-A+i\sigma}^{+A+i\sigma}F(\Omega)\,{\rm
    e}^{-i\,\Omega\tau}d\Omega \,,
\end{equation}
where a real positive parameter $\sigma$ exceeds the so-called
growth index $\sigma_0>0$ of the original function $f(\tau)$. The
Laplace image $F(\Omega)$, as a function of the complex variable
$\Omega{=}\omega {+}i \gamma$, is defined and analytic in the
domain $\Im \Omega {=} \gamma>\sigma$  of the plane $\omega 0
\gamma$. Second, in many interesting cases, two points $\Omega {=}
\pm k$ happen to be branchpoints of the function $F(\Omega)$ (see,
e.g., \cite{Silin,2} for details); as we will show below, in our
case this rule remains valid. Third, in order to use the theorem
about residues, we should prolong the integration contour into the
domain $\Im \Omega < \sigma$, harboring all the poles of the
function $F(\Omega)$ and keeping in mind that the branchpoints
have to remain the external ones. We use the contour presented on
Fig.~\ref{Contur}, the radius of the arc being
$R{=}\sqrt{A^2{+}\sigma^2}$.

In order to apply the residues theorem to the calculation of the
function $F_{ik}(\tau, x^{\alpha})$ we should fix one of the
analytic continuations of the function ${\cal F}_{ik}(\Omega,
k_{\alpha})$ into the domain $\Im \Omega < \sigma$. Thus, the
function ${\cal F}_{ik}(\tau, x^{\alpha})$ contains contributions
of two types: first, the residues in the poles of the function
${\cal F}_{lk}(\Omega,k_\gamma)$; second, the integrals along the
cuts $\omega {=} \pm k$, $\gamma {=} k y$, ${-}\infty<y \leq0$.
The contribution of the second type displays the dependence on
time in the form $\exp{(ik \tau)}$, and looks like the packet of
waves propagating with the phase velocity $V_{{\rm ph}}{=} c{=}1$.
In order to describe the contributions of the first type, below we
study in detail the solutions to the dispersion relations for the
longitudinal and transversal waves in plasma. Providing the
mentioned analytic continuation of the function ${\cal
F}_{lk}(\Omega,k_\gamma)$ into the domain $\Im \Omega < \sigma$,
we need to take special attention to the analytic properties of
the permittivity scalar $\varepsilon_{||}$ (see (\ref{e4})), which
is one of the structural elements of the longitudinal electric
field. The discussion of this problem started in \cite{Vlasov1}
and led to the appearance in the scientific lexicon of the term
Landau damping \cite{Silin,2,LLX,4}, based on the prediction made
in \cite{Landau}. The results of this discussion briefly can be
formulated as follows (we assume here that the growth index
vanishes, i.e., $\sigma{=}0$). The most important part of the
integral (\ref{e4}) is the integration with respect to the
longitudinal velocity $v_{||} {=}
\frac{k_{\alpha}q^{\alpha}}{kq}$, which can be written as
$\int_{{-}1}^{{+}1} \frac{dv_{||}
Z(v_{||})}{\left(v_{||}{-}\frac{\Omega}{k}\right)}$. When $\Im
\Omega {=}\gamma{=}0$,  this real integral diverges at $\omega<k$,
and the function $\varepsilon_{||}$ is not defined. When $\gamma
\neq 0$,  this integral can be rewritten as a contour integral
with respect to complex velocity $v_{||}{=}x{+} iy$. Since in the
domain $\gamma>0$ the function should be analytic, and the pole
$v_{||}{=}\frac{\Omega}{k}$ can appear in the lower semi-plane
$\gamma<0$ only, we recover (in our terminology) the classical
Landau's statement about resonant damping of the longitudinal
waves in plasma, which can take place if the plasma particles
co-move with the plasma wave and extract the energy from the
plasma wave \cite{Landau}.

\section{Analysis of the dispersion relations. I. Longitudinal waves}\label{sec4}

\subsection{Dispersion equation for the ultrarelativistic plasma}

Below we consider the background state of the ultrarelativistic plasma to be described by the distribution functions
\begin{equation}
f^{(0)}_{({\rm a})}(q) = \frac{N_{({\rm a})}}{8\pi T_{({\rm
a})}^3}\,{\rm e}^{-\frac{q}{T_{({\rm a})}}} \,,
\end{equation}
where the temperatures for all sorts of particles should coincide
if the background state was the equilibrium one. We are
interested in the analysis of the solutions
$\Omega(k_\alpha,\nu){=}\omega{+}i \gamma$ of the dispersion relations
for longitudinal (\ref{longDR}) and transversal (\ref{transDR})
waves. More precisely, we focus on the classification of the roots
of (\ref{longDR}) and (\ref{transDR}) and search for nonstandard
solutions which appear just due to nonminimal interactions and
axion-photon couplings. A number of facts, which we discuss below,
are well-known in the context of relativistic plasma-wave theory;
nevertheless, we prefer to restate them in order to explain
properly new results, which appear in the axionically active
plasma nonminimally coupled to gravity.
We have to stress that in the model under consideration the results
of integration are presented in an explicit form, and the analytic
continuation of all the necessary functions also is made explicitly.

Since the background state of plasma is spatially isotropic, one can
choose the $0Z$ axis along the wave vector, i.e., without loss of generality one can put
\begin{equation}
k_\alpha=(0,0,-k)\,,\quad q_\alpha = q\,(\cos\varphi\sin\theta,
\sin\varphi\sin\theta, \cos\theta)\,. \label{an1}
\end{equation}
Then the longitudinal permittivity scalar (\ref{e4}) can be
reduced to the following term:
\begin{equation}
    \eps_{||}= 1- \frac{3W^2}{k^2} \ \textsf{Q}_1(z) \,,\label{an2}
\end{equation}
where $t=\cos\theta$, $z \equiv {\Omega}/{k}$, the term
\begin{equation}
W^2 = \frac{4\pi}{3} \sum_{({\rm a})} \frac{e^2_{({\rm a})} N_{({\rm a})}}{T_{({\rm a})}}
\label{an3}
\end{equation}
is usually associated with the square of the plasma frequency in
the ultrarelativistic approximation \cite{Silin}, and, finally,
the Legendre function of the second kind $\textsf{Q}_1(z)$ is
given by the integral (see, e.g., \cite{Hobson,Bateman})
\begin{equation}
\textsf{Q}_1(z) = \frac12  \int\limits_{-1}^{1}\frac{t \,dt}{z-t}
= \Real \textsf{Q}_1(z) + i \Im \textsf{Q}_1(z) \,,\label{an4}
\end{equation}
with
\begin{gather}
\Real \textsf{Q}_1(z) = -1 + \frac{x}{4} \log
{\left[\frac{(x+1)^2+y^2}{(x-1)^2+y^2}\right]} - {}\nonumber\\
{}-\frac{y}{2}
\left[\arctan{\frac{x-1}{y}}-\arctan{\frac{x{+}1}{y}} \right]
\,,\label{an6}
\end{gather}
\begin{gather}
\Im \textsf{Q}_1(z) = \frac{y}{4} \log
{\left[\frac{(x+1)^2+y^2}{(x-1)^2+y^2}\right]} +{}\nonumber\\ {}+
\frac{x}{2} \left[\arctan{\frac{x-1}{y}}-\arctan{\frac{x+1}{y}}
\right] \,.\label{an7}
\end{gather}
We treat the quantity $z=\frac{\Omega}{k}$ as a new complex variable
$z=x+iy$, where $x=\frac{\omega}{k}$ and $y=\frac{\gamma}{k}$. In terms of
the complex variable $z$ this function looks more attractive
\begin{equation}
\textsf{Q}_1(z) = \frac12\,z\ln \left(\frac{z+1}{z-1}\right)-1
\,,  \quad \textsf{Q}_1(0) = -1 \,, \label{an8}
\end{equation}
nevertheless, one should, as usual, clarify analytical properties
of this function. Clearly, the points $z=\pm 1$ are the
logarithmic branchpoints of the function  $\textsf{Q}_1(z)$; in these two
points the real part of the Legendre function does not exist. When
we cross the line $\Im z=0$ on the fragment $|\Real z|<1$ of the
real axis, the function $\Im \textsf{Q}_1(z)$ experiences the
jump, since
\begin{gather}
\lim_{y \to 0_{{+}0}}\{\Im \textsf{Q}_1 \}=- x \frac{\pi}{2}
\,,\nonumber \\ \lim_{y \to 0_{{-}0}}\{\Im \textsf{Q}_1 \}=+ x
\frac{\pi}{2} \,, \quad |x|<1 \,. \label{an9}
\end{gather}
When $|x|>1$, the function $\Im \textsf{Q}_1(z)$ is continuous. In
order to obtain analytical function $\eps_{||}(z)$ we consider the
function $\textsf{Q}_1(z)$ to be defined in the domain $\Im z>0$
and make the analytical extension to the domain $\Im z<0$ as
follows:
\begin{gather}
\textsf{Q}_1(z) \to \textsf{G}_{||}(z) \equiv
\frac12\int\limits_{-1}^{1}\frac{t \,dt}{z-t} -{}\nonumber\\{}-
i\pi z\,\Theta(-\Im z)\Theta(1-|\Real z|)\,. \label{Glong}
\end{gather}
Here $\Theta$ denotes the Heaviside step function. Let us mention
that the function $\textsf{G}_{||}(z)$ is analytic everywhere
except the lines $z{=}\pm1{+}iy$ with ${-}\infty<y\leq0$. In fact, to
obtain the analytic continuation, we added the residue $({-}i\pi z)$
in the singular point $t{=}z$ into the Legendre function. The same
result can be obtained if one deforms the integration contour in
(\ref{an4}) so that this contour lies below the singular point
$t=z$ and harbors it; in the last case we would repeat the method
applied by Landau in \cite{Landau}.

Let us note that the function $\textsf{G}_{||}(z)$ possesses the symmetry
\begin{equation}\label{Gconj}
    \overline{\textsf{G}_{||}(z)}=\textsf{G}_{||}(-\bar{z})\,,
    \end{equation}
i.e., it keeps the form with the transformation
$z\to -\bar{z}$, which is equivalent to $\omega\to -\omega$.
Keeping in mind this fact, below we consider $\omega$ to be nonnegative without loss of generality.

Now the function
\begin{equation}
\eps_{||}= 1 - \frac{3W^2}{k^2}\,\textsf{G}_{||}(z)
\label{epslong2}
\end{equation}
is defined and is analytical on the complex plane $z$ everywhere
except the branchpoints $z{=}\pm 1$. The corresponding dispersion
relation for longitudinal plasma waves can be written as follows:
\begin{equation}
\frac{k^2(1+2K_1)}{3W^2} = \textsf{G}_{||}(z) \,. \label{longDR21}
\end{equation}

When $K_1{=}0$, this equation gives the well-known results (see,
e.g., \cite{Silin}). When $K_1 \neq 0$, the results are obtained
in \cite{BM09} for three different cases: $1+2K_1>0$,  $1+2K_1<0$
and $1+2K_1=0$. We recover these results only to demonstrate the
method, which we use below for the dispersion equations containing
the axionic factor $\nu \neq 0$. Since the left-hand side of
Eq.(\ref{longDR21}) is real, we require that
\begin{equation}
\frac{k^2(1+2K_1)}{3W^2}   =  \Real \{\textsf{G}_{||}(z)\} \,,
\quad \Im \{\textsf{G}_{||}(z)\}{=}0 \,, \label{52longDR}
\end{equation}
or in more detail
\begin{widetext}
\begin{equation}
\frac{x}{4} \log {\left[\frac{(x+1)^2+y^2}{(x-1)^2+y^2}\right]} -
\frac{y}{2} \left[\arctan{\frac{x-1}{y}}-\arctan{\frac{x+1}{y}}
\right]+ \pi y\, \Theta(-y) \Theta(1-|x|) = \mu , \label{an69}
\end{equation}
\begin{equation}
\frac{y}{4} \log {\left[\frac{(x+1)^2+y^2}{(x-1)^2+y^2}\right]} +
\frac{x}{2} \left[\arctan{\frac{x-1}{y}}-\arctan{\frac{x+1}{y}}
\right] - \pi x\, \Theta(-y) \Theta(1-|x|) = 0,\label{an77}
\end{equation}
\end{widetext}
where we introduced the auxiliary parameter $\mu$
\begin{equation}
\mu \equiv 1+ \frac{k^2(1+2K_1)}{3W^2} \,, \label{59longDR}
\end{equation}
which is the function of $k$.
There are two explicit solutions to these equations.

\subsection{Solutions with $x=0$}

When $x{=}\frac{\omega}{k}{=}0$, the equation (\ref{an77}) is satisfied
identically, and we can assume that either the frequency vanishes
$\omega{=}0$, or the wavelength becomes infinitely small, $k{=}\infty$. The equation
(\ref{an69}) can be written now as
\begin{equation}
\mu= y\left(\arctan{\frac1y} {+} \pi \Theta (-y)\right)\quad
\Leftrightarrow \quad \arccot y = \frac{\mu}{y}\,.
\label{2longDR27}
\end{equation}
Clearly, when  $\mu>1$ (or equivalently, $1{+}2K_1 > 0$) there are
no solutions to this equation. When $1{+}2K_1 < 0$, i.e., $\mu
<1$, one solution $y=y^*$ to the equation (\ref{2longDR27})
appears: it is positive if $k^2< \frac{3W^2}{|1{+}2K_1|}$ and
negative for $k^2> \frac{3W^2}{|1{+}2K_1|}$. In the intermediate
case $\mu{=}1$, i.e.,  when $1{+}2K_1 {=} 0$, we obtain formally
the solution $y{=}\infty$, which we omit as nonphysical.

\subsection{Solutions with $y=0$ and $|x|>1$}

When $y=0$, we deal with waves propagating without
damping/increasing, since $\gamma =0$. The frequency $\omega$ can
be now found from the equation
\begin{equation}
\mu = \frac{\omega}{2k} \log{\left(\frac{\omega + k}{\omega -
k}\right)} \,. \label{longDR30}
\end{equation}
In terms of $x$ and $\mu$ introduced in (\ref{59longDR}) this
equation can be rewritten as $\frac{1}{x} {=}\tanh{\frac{\mu}{x}}$, thus we obtain
the following results.

First, there are no solutions, when $\mu
\leq 1$, i.e., when $1{+}2K_1 \leq 0$. Second, there are one
positive and one negative roots ($x=\pm x^*$), when $\mu >1$, i.e., when $1{+}2K_1$ is positive.

\begin{figure}[h]
\includegraphics[width=4.5cm]{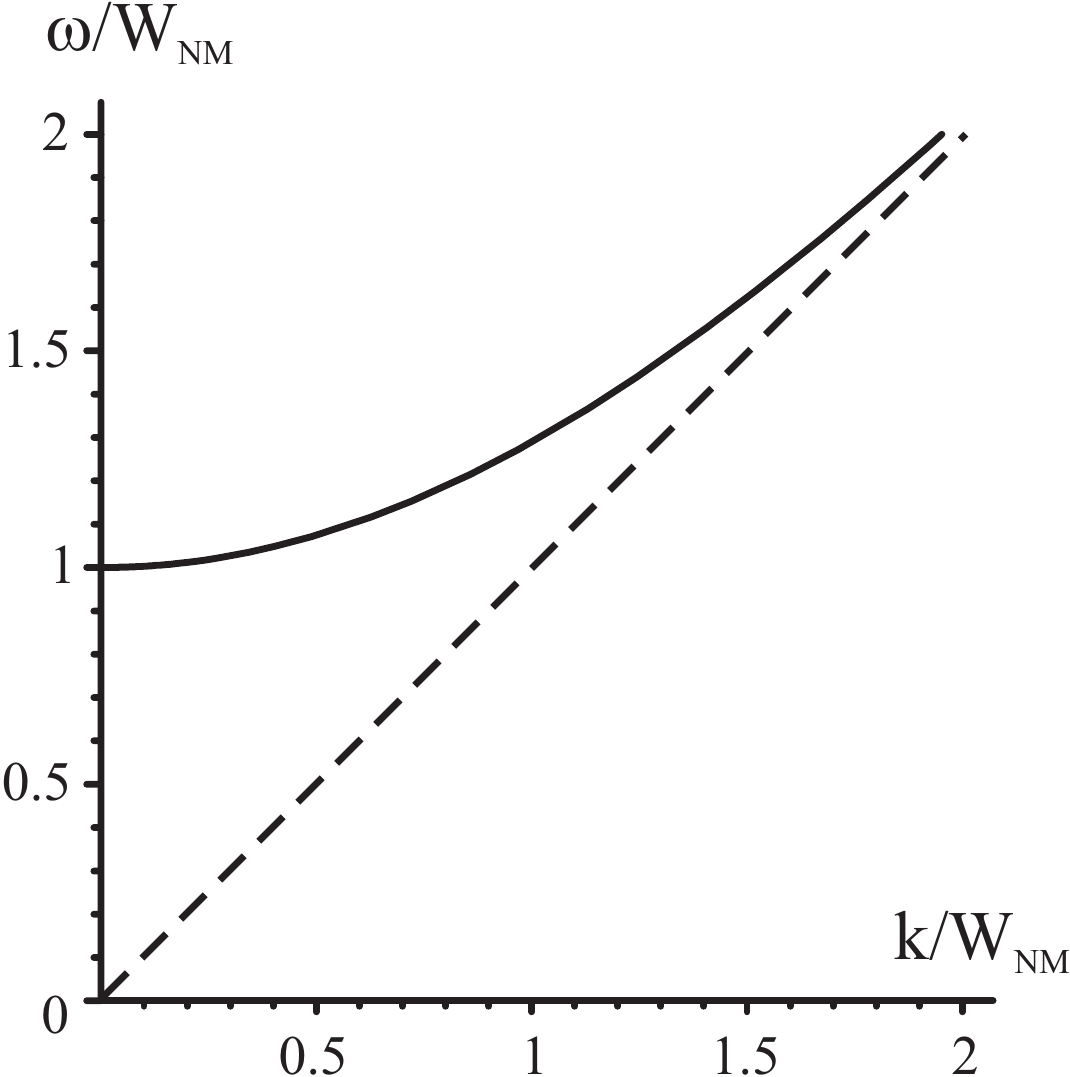}\\ \includegraphics[width=5cm]{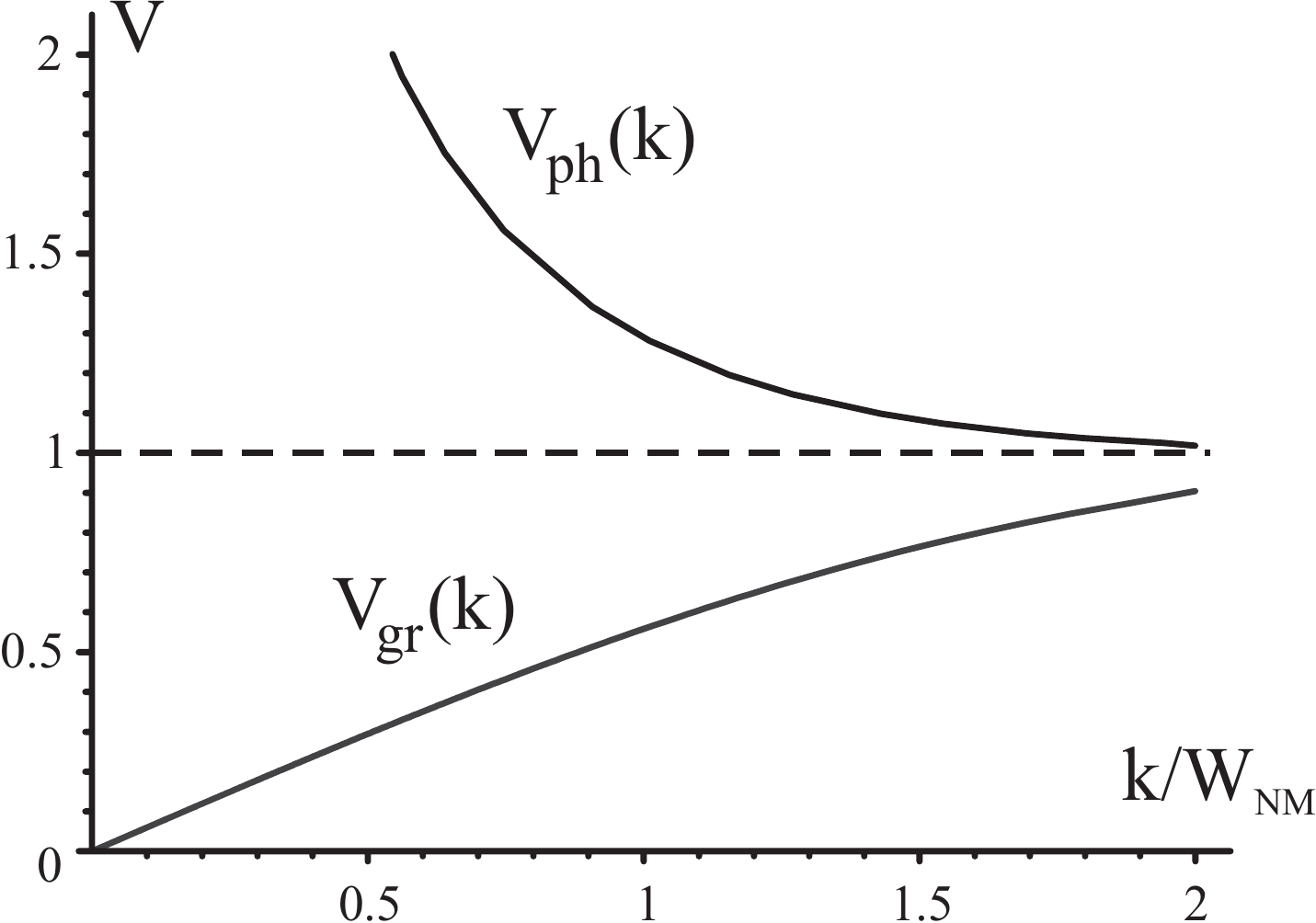}
\caption{The plots of functions $\omega(k)$, $V_{{\rm ph}}(k)$ and
$V_{{\rm gr}}(k)$ for running longitudinal waves in the
ultrarelativistic plasma in the case, when $1{+}2K_1>0$. The phase
velocity exceeds the speed of light in vacuum, $V_{{\rm
ph}}(k)>1$, while the group velocity is less than one, $V_{{\rm
gr}}(k)<1$ for arbitrary $k$.} \label{wlong}
\end{figure}

To estimate the interval of frequencies, which are allowed by the
equation (\ref{longDR30}),  one can use two decompositions for the
Legendre function in the case, when $y=0$ and $x>1$:
\begin{gather}
{\textsf Q}_1(x)=-1 +
\frac{x}{2}\log{\left(\frac{x+1}{x-1}\right)} =
\sum_{n=1}^{\infty}\frac{x^{-2n}}{2n+1}={}\nonumber\\{}=\frac{1}{3x^2}+\frac{1}{5x^4}+\dots
\,,\label{expan1}
\end{gather}
\begin{equation}
(x^2-1){\textsf
Q}_1(x)=\frac13-2\sum_{n=1}^{\infty}\frac{x^{-2n}}{(2n+1)(2n+3)} \,.\label{expan2}
\end{equation}
This means that $\frac{1}{3x^2}<{\textsf
Q}_1(x)<\frac{1}{3(x^2-1)}$ and thus, taking into account that
${\textsf Q}_1(x)=\frac{k^2}{3W_{{\rm NM}}^2}$ (see
(\ref{longDR21})), we obtain the inequality $W_{{\rm
NM}}<\omega<\sqrt{W_{{\rm NM}}^2+k^2}$. In particular, when $k\ll
\omega$, one obtains from (\ref{longDR30})
\begin{equation}
\omega^2 \simeq W_{{\rm NM}}^2+ \frac{3}{5}k^2 \,,
\label{longDR31}
\end{equation}
and this result is well-known at $K_1{=}0$ (see, e.g.,
\cite{LLX,Silin}). Here and below we use the convenient parameter
$W_{{\rm NM}} \equiv \frac{W}{\sqrt{|1{+}2K_1|}}$. The plots of
the functions $\omega(k)$, $V_{{\rm ph}}(k)$ and $V_{{\rm gr}}(k)$
are presented on Fig.~\ref{wlong}.

Let us remark, that depending on the value of the wave parameter
$k$ two types of longitudinal waves in the relativistic plasma can
exist (see, e.g., \cite{Silin}). The waves of the first type
propagate without damping ($\gamma{=}0$) and with the phase
velocity exceeding speed of light in vacuum $V_{\rm ph}(k)>1$; for
instance, the well-known Langmuir waves are described by the
dispersion relation $\omega^2{=}\omega^2_{({\rm p})}{+}k^2
V^2_{({\rm s})}$, thus, $V_{\rm ph}(k)>1$, when
$k<\frac{\omega_{({\rm p})}}{\sqrt{1{-}V^2_{({\rm s})}}}$ (here
$\omega_{({\rm p})}$ is the Langmuir plasma frequency, and
$V_{({\rm s})}$ is the reduced sound velocity). The longitudinal
plasma waves of the second type are characterized by $\gamma<0$
and $V_{\rm ph}<1$, thus displaying the well-known Landau damping
phenomenon. However, in the ultrarelativistic limit, as it was
shown in \cite{Silin}, the plasma waves of the second type are
suppressed, and only the waves of the first type exist. In fact,
we have shown here that the account for a nonminimal coupling of
the electromagnetic field to gravity does not change this result
for ultrarelativistic case, if $1{+}2K_1>0$. In the case, when
$1{+}2K_1 \leq 0$, i.e., when the curvature induced effects are
not negligible, the longitudinal waves of the first type in the
ultrarelativistic plasma are also suppressed, i.e., no
longitudinal waves in the ultrarelativistic plasma can be
generated, whatever the value of $k$ is chosen, if $1{+}2K_1 \leq
0$.

\subsection{About solutions with $y \neq 0$ and $x \neq 0$}

The question arises, whether the longitudinal wave exists for
which both $x$ and $y$ are nonvanishing? In order to answer this
question let us transform the pair of equations (\ref{an69}),
(\ref{an77}) into
\begin{equation}
\mu = \frac{(x^2+y^2)}{4 x}
\log{\left[\frac{(x+1)^2+y^2}{(x-1)^2+y^2}\right]} \,,
\label{an79}
\end{equation}
\begin{gather}
\frac{2\mu y}{(x^2+y^2)} +
\left[\arctan{\frac{x-1}{y}}-\arctan{\frac{x+1}{y}} \right]
-{}\nonumber\\{}- 2\pi \Theta(-y) \Theta(1-|x|) =0 \,,
\label{an87}
\end{gather}
assuming that $y \neq 0$ and $x \neq 0$. Since the expression in
the right-hand side of (\ref{an79}) is positive, we should
consider the situations with $\mu>0$ only. We have to distinguish
two important cases.

\subsubsection{$y<0$ and $x^2<1$}

\noindent
In this case the equation (\ref{an87}) takes the form
\begin{equation}
\frac{2\mu |y|}{x^2+y^2} +
 \left[2\pi + \arctan{\frac{x-1}{|y|}}-\arctan{\frac{x+1}{|y|}} \right]  =0 \,. \label{an871}
\end{equation}
Since the function $\arctan{U}$ satisfies the condition $|\arctan{U}| \leq \frac{\pi}{2}$, both the first term and the term in brackets are positive,
thus guaranteeing that this equation has no roots in this case.

\subsubsection{$y>0$ or/and $x^2>1$}

\noindent In this case we return to the Legendre function and
transform its imaginary part as follows
\begin{gather}
\Im\textsf{Q}_1(z)=\frac{1}{2i}\left(\textsf{Q}_1(z)-\textsf{Q}_1(\bar{z})\right)={}\nonumber\\
{}=\frac{1}{4i}\int\limits_{-1}^1\frac{(\bar{z}-z)\,t\,dt}{|z-t|^2}
    =-\frac12\, y \,\int\limits_{-1}^1\frac{t\,dt}{|z-t|^2}
    ={}\nonumber\\
{}=-\frac14\, y\,
\int\limits_{-1}^1t\,dt\,\left(\frac{1}{|z-t|^2}-\frac{1}{|z+t|^2}\right)={}\nonumber\\
   {} =- x y \int\limits_{-1}^1\frac{t^2\,dt}{|z^2-t^2|^2} \,.
\label{ImQ}
\end{gather}
Clearly, the integral in (\ref{ImQ}) is a positive quantity, thus,
$\Im\textsf{Q}_1(z)=0$ if and only if $x \cdot y=0$. In other
words, there are no solutions of the dispersion relation under
discussion, when both $x$ and $y$ are nonvanishing.

\section{Analysis of the dispersion relations. II. Transversal waves}\label{sec5}

\subsection{Dispersion equation}

In analogy with the longitudinal permittivity scalar, the
transversal one can be written in terms of reduced frequency $z
=\frac{(\omega {+}i \gamma)}{k}$, of the wave three-vector modulus $k$ and
the quantity $W$ (see (\ref{an3})) only:
\begin{equation}
\eps_{\bot}= -\frac{3W^2}{2z^2k^2}\,\textsf{G}_{\bot}(z)\,,
\label{epstrans1}
\end{equation}
where
\begin{equation}
\textsf{G}_{\bot}(z) = \left[1-(z^2-1)\textsf{G}_{||}(z)\right] \,, \quad \textsf{G}_{\bot}(0) = 0 \,.
\label{epstrans2}
\end{equation}
Here we used the definition (\ref{e3}) of the scalar $\eps_{\bot}$
and the formulas (\ref{e1})-(\ref{e4}). The function
$\textsf{G}_{\bot}(z)$ also possesses the property
\begin{equation}\label{Gconj3}
\overline{\textsf{G}_{\bot}(z)}=\textsf{G}_{\bot}(-\bar{z}) \,,
\end{equation}
and is analytic with respect to complex variable $z$ anywhere except the lines $z=\pm1+iy$ with $-\infty<y\leq0$. In
these terms the dispersion relation for transversal waves in
ultrarelativistic plasma takes the form
\begin{equation}
k^2(z^2-1) + k p - \frac{3}{2} W_{{\rm NM}}^2 \textsf{G}_{\bot}(z)
\sgn(1+2K_1) =0 \,, \label{transDR13}
\end{equation}
where the real parameter
\begin{equation}
    p= \mp \nu \frac{(1+2K_2)}{(1+2K_1)}
\label{p1}
\end{equation}
appears if and only if the axion field is nonstationary. Since
$\textsf{G}_{\bot}(0){=}0$, the solution $z{=}0$ exists, when
$k{=}p \geq 0$. Depending on the value of the parameter
$(1{+}2K_1)$ one obtains three particular cases.

\subsection{The first case: $1+2K_1>0$ and $x^2>1$ or/and $y>0$}

First, we consider the imaginary part of the equality
(\ref{transDR13}) along the line the analysis for the longitudinal
case. When $\textsf{G}_{||}(z){=} \textsf{Q}_{1}(z)$, i.e. when
either $\Theta({-}y){=}0$ or $\Theta(1{-}|x|){=}0$, we can show explicitly
that
$$
\Im\{k^2(z^2-1) + k p-\frac{3}{2} W_{{\rm NM}}^2
\textsf{G}_{\bot}(z)\} =
$$
\begin{equation}
= x \cdot y \left\{2k^2+\frac{3W_{{\rm
NM}}^2}{2}\int\limits_{-1}^1\frac{t^2\left(1-t^2\right)\,dt}{|z^2-t^2|^2}\right\}
\,. \label{tr45}
\end{equation}
Clearly, this expression can be equal to zero if and only if $x=0$
or $y=0$.

\subsubsection{On the solutions with $y=0$ and  $x>1$}

\begin{figure}[h]
\includegraphics[width=8cm]{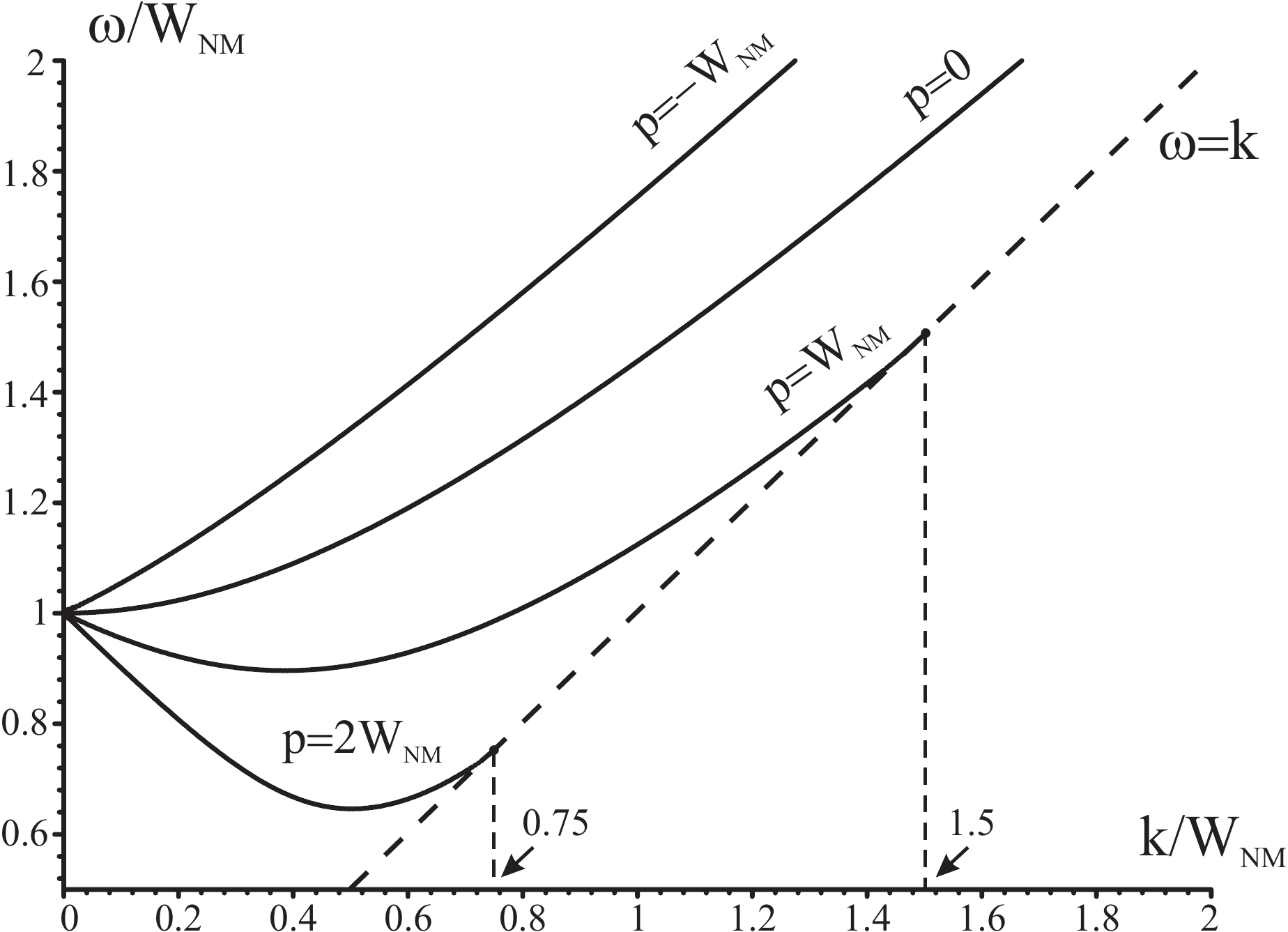}
\caption{Plots of the dispersion function $\omega(k)$ for running
transversal waves with $\gamma{=}0$ and $\omega>k$ at
$1{+}2K_1>0$. When $p>0$ the lines $\omega{=}\omega(k)$ stop on
the bisector line $\omega{=}k$ at $k=k_{{\rm crit}}{=}\frac{3
W_{{\rm NM}}^2}{2p}$, and cannot be prolonged for $k>k_{{\rm
crit}}$. For negative values of $p$ the curves do not cross and do
not touch the bisector $\omega {=}k$ at $k \to \infty$; when
$p{=}0$, the bisector is the asymptote for the corresponding
curve.} \label{wtrans}
\end{figure}

\begin{figure}[h]
\includegraphics[height=5cm]{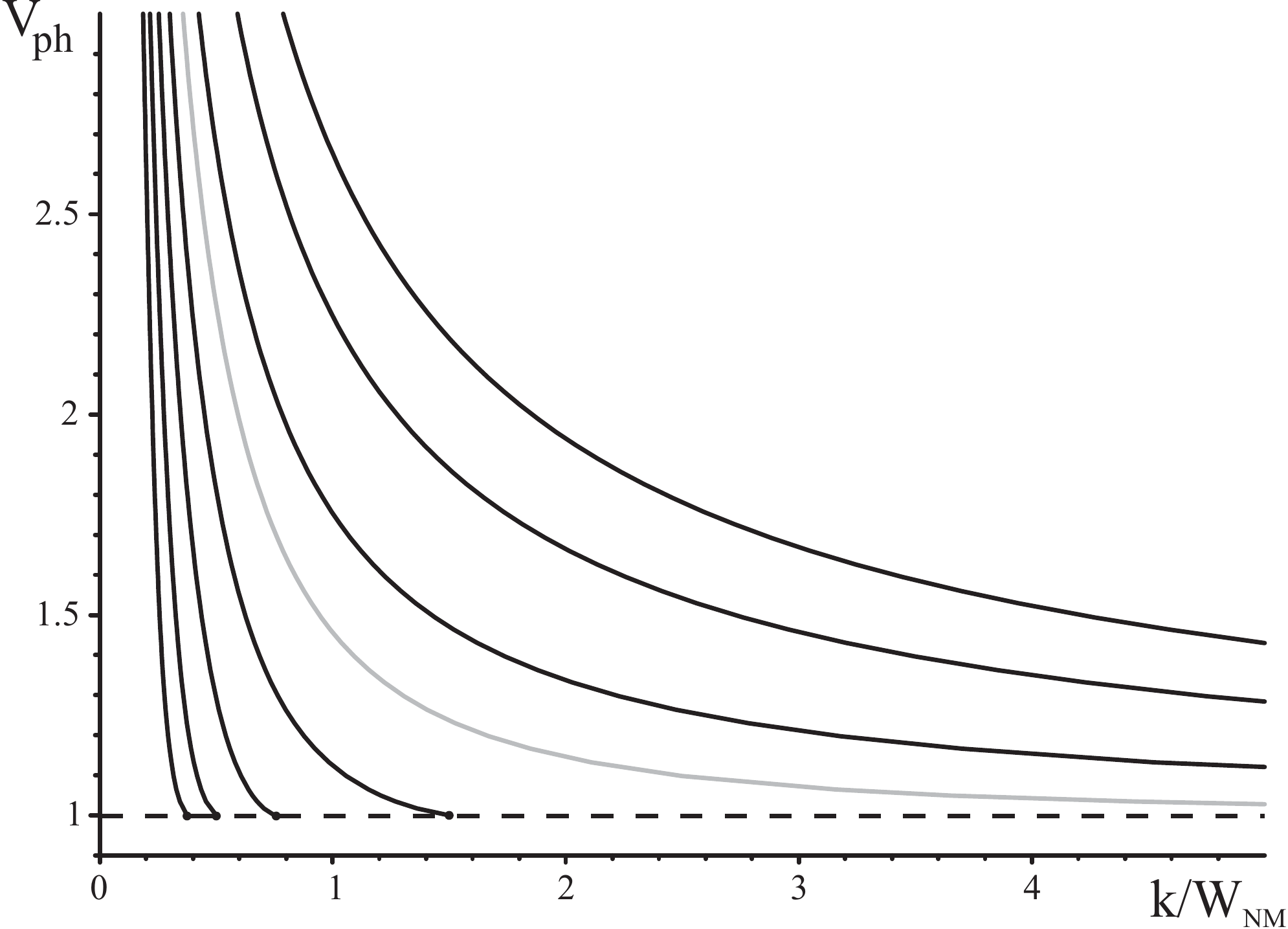}\\
\includegraphics[height=5cm]{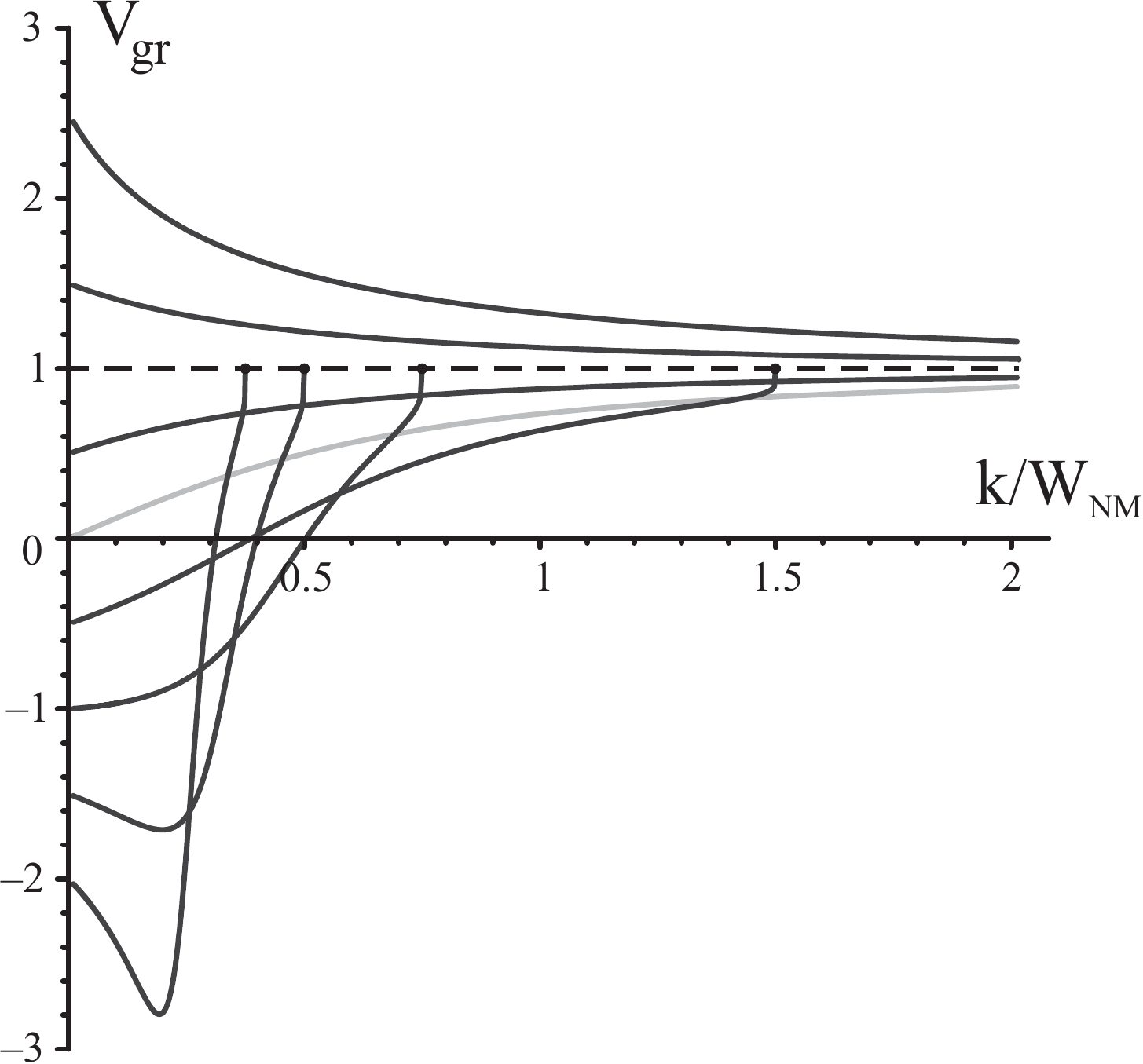}
\caption{Plots of the phase and group velocities for running
transversal waves with $\gamma{=}0$ and $\omega>k$ at
$1{+}2K_1>0$. When $p>0$, both $V_{{\rm ph}}$ and $V_{{\rm gr}}$
become equal to the speed of light in vacuum at $k=k_{{\rm
crit}}{=}\frac{3 W_{{\rm NM}}^2}{2p}$; the solutions of the
dispersion equation do not exist for $k>k_{{\rm crit}}$. When
$p{=} 0$ (light grey curves),  the straight line $V{=}1$ is the
horizontal asymptote for curves describing both  phase and group
velocities at $k \to \infty$. When $p$ is negative, and
$|p|<p_{{\rm crit}}$, one sees that $V_{{\rm gr}}(k)<1$ for
arbitrary $k$; when $p<0$ and $|p|>p_{{\rm crit}}$, the group
velocity exceeds the speed of light in vacuum $V_{{\rm gr}}>1$.}
\label{vph-trans+}
\end{figure}

\noindent When $y=0$ the dispersion relation gives
\begin{equation}
k=\frac{-p \pm \sqrt{p^2+6W_{{\rm
NM}}^2(x^2-1)\textsf{G}_\bot(x)}}{2(x^2-1)} \,, \label{p14}
\end{equation}
where the function $\textsf{G}_\bot(x)$ can be written as
\begin{equation}
\textsf{G}_\bot(x) = x^2\left[1- \frac{(x^2-1)}{2x}
\log{\left(\frac{x+1}{x-1}\right)} \right]\,. \label{p25}
\end{equation}
For this function the following decompositions can be found:
\begin{equation}
{\textsf G}_\bot(x)=1-(x^2-1){\textsf
Q}_1(x)=2\sum_{n=0}^{\infty}\frac{x^{-2n}}{(2n+1)(2n+3)}\,,\label{expan3}
\end{equation}
\begin{equation}
(x^2-1){\textsf
G}_\bot(x)=\frac{2x^2}{3}-8\sum_{n=0}^{\infty}\frac{x^{-2n}}{(2n+1)(2n+3)(2n+5)}\,,\label{expan4}
\end{equation}
from which one can see that $\textsf{G}_\bot(1)=1$ and
$\textsf{G}_\bot(\infty)=\frac{2}{3}$. Being monotonic, the
function $\textsf{G}_\bot(x)$ is restricted by the inequalities
$\frac{2}{3}<\textsf{G}_\bot(x)<1$, when $x>1$. This means that
the square roots in equations (\ref{p14}) is real, the right-hand
side is positive for arbitrary $p$, if we use only the root with
plus sign in front of the square root. Thus, the equation
(\ref{p14}) (with the plus sign in front of the square root) has
real solutions $\Omega =\omega(k)$, which describe transversal
oscillations without damping/increasing and phase velocity
exceeding the speed of light in vacuum. When $p$ is nonpositive,
these oscillations have arbitrary wavelength
$0<\frac{1}{k}<\infty$. When $p$ is positive, $k$ cannot exceed
some critical value $k_{{\rm crit}} \equiv \frac{3 W_{{\rm
NM}}^2}{2p}$, which can be obtained from the following estimation:
\begin{gather}
k=\frac{3W_{{\rm NM}}^2 \textsf{G}_\bot(x)}{p + \sqrt{p^2+6W_{{\rm
NM}}^2(x^2-1)\textsf{G}_\bot(x)}}
\nonumber\\
{}< \frac{3W_{{\rm NM}}^2 \textsf{G}_\bot(x)}{2p} < \frac{3W_{{\rm
NM}}^2}{2p}\,. \label{p141}
\end{gather}
In the approximation of long waves, i.e., when $k \to 0$, we obtain
\begin{equation}
\omega^2(k) \simeq W^2_{{\rm
NM}}-p\,k+\frac65\,k^2+\frac{p}{5W_{\rm NM}^2}\,k^3 \,.
\label{appr1}
\end{equation}
For short waves ($k \to \infty$) and $p\leq 0$ the square of the frequency reads
\begin{equation}
\omega^2(k) \simeq k^2 - p\,k + \frac{3}{2} W^2_{{\rm NM}} \,.
\label{appr2}
\end{equation}
In Fig.~\ref{wtrans} and Fig.~\ref{vph-trans+}  we presented the
results of numerical calculations: we have chosen four values of
the parameter $p$ for illustration of the dependence $\omega(k)$,
and eight values of $p$ for illustration of the behavior of the
functions $V_{{\rm ph}}(k)$ and $V_{{\rm gr}}(k)$.

Let us remark that, when the parameter $p$ is negative, there
exists some critical value of its modulus $|p|{=}p_{{\rm crit}}$,
which distinguishes two principally different situations: when
$|p|<p_{{\rm crit}}$, the group velocity does not exceed the speed
of light in vacuum $V_{{\rm gr}}<1$;  when $|p|>p_{{\rm crit}}$,
we obtain $V_{{\rm gr}}>1$ for arbitrary $k$. This critical
behavior of the function $V_{{\rm gr}}(k)$ can be easily
illustrated for the case of short waves (see (\ref{appr2})).
Indeed, in this case we obtain
\begin{equation}
V_{{\rm gr}} = \frac{d\omega}{dk}\simeq \left[1 +
\frac{(6W^2_{{\rm NM}}-|p|^2)}{(2k+|p|)^2} \right]^{-\frac12} \,,
\label{appr29}
\end{equation}
and, clearly, $p_{{\rm crit}} {=} \sqrt6  W_{{\rm NM}}$. Another
interesting feature is that the function $V_{{\rm gr}}(k)$ can
take negative values for some values of the parameter $p$. When we
think about both results: first, that $V_{{\rm gr}}(k)$ is
negative for some $p$, and that $V_{{\rm gr}}(k)$ can exceed the
speed of light in vacuum, $V_{{\rm gr}}>1$, we recall the theory
of magneto-active plasma, which demonstrates such behavior for the
waves of special type (see, e.g., \cite{GR}). It can be considered
as a supplementary motif to use the term axionically active plasma
by the analogy with the term magneto-active plasma.

\subsubsection{On the solutions with $x=0$ and $y>0$}

When $x=0$ and $y>0$, the solution to the dispersion equation is
\begin{equation}
k=\frac{p \pm \sqrt{p^2-6W_{{\rm
NM}}^2(y^2+1)\textsf{G}_\bot(iy)}}{2(y^2+1)} \,, \label{p34}
\end{equation}
where
\begin{equation}
\textsf{G}_\bot(iy) =  y^2\left[\frac{(y^2+1)}{y}
\arctan{\frac{1}{y}} - 1\right]\,. \label{p35}
\end{equation}
For positive values of the variable $y$ the function
$\textsf{G}_\bot(iy)$ is nonnegative
($\textsf{G}_{\bot}(0_{+0})=0$, $\textsf{G}_{\bot}(+i\infty) =
{2}/{3}$). This means that, first, at $p=0$ there are no solutions
to (\ref{p34}); second, at $p \neq 0$ the corresponding solutions
exist, for which $p^2 \geq 6W_{{\rm
NM}}^2(y^2+1)\textsf{G}_\bot(iy)$; third, when $p>0$ one can use
both signs, plus and minus, in (\ref{p34}); fourth, when $p<0$,
there are no solutions, since $k$ should be positive. Let us
stress the following feature: since $\textsf{G}_\bot(iy)>0$ we
obtain from (\ref{p34}) the inequality $k<p$ for solutions with
$x=0$ and $y>0$. To conclude, the nonharmonically increasing
perturbations exist ($\Omega  = i \gamma$ with $\gamma>0$) only in
case, when $p>k$.

\subsection{The second case: $1+2K_1>0$ and $x^2<1$, $y<0$}

When $y<0$ and $|x|<1$, we divide the equation (\ref{transDR13})
by the quantity $(z^2{-}1)$ and transform the imaginary part of obtained relation
into
\begin{gather}
\left(kp-\frac{3W_{{\rm NM}}^2}{2}\right) \frac{2x|y|}{|z^2-1|^2}
+{}\nonumber\\{}+ \frac{3W_{{\rm NM}}^2}{2}\left[\Im
{\textsf{Q}}_{1}(z)-\pi x \right]=0 \,. \label{tr46}
\end{gather}
Keeping in mind the properties of the function $\Im
{\textsf{Q}}_{1}(z)$, which we studied for the case of
longitudinal waves, one can state that, when $p \leq
\frac{3W_{{\rm NM}}^2}{2k}$, the left-hand side of the equation
(\ref{tr46}) is nonpositive, and thus there is only one root of
this equation, namely, $x=0$. This case can be considered
qualitatively. When $p > \frac{3W_{{\rm NM}}^2}{2k}$, we will
study solutions numerically.

\subsubsection{The case $x=0$ and $y<0$}

Now the dispersion equation yields
\begin{equation}
k=\frac{p \pm \sqrt{p^2+6W_{{\rm
NM}}^2(|y|^2+1)\left[-\textsf{G}_\bot(-i|y|)\right]}}{2(|y|^2+1)}
\,, \label{p44}
\end{equation}
where the function
\begin{equation}
\left[-\textsf{G}_\bot(-i|y|)\right] =  |y|^2\left[1+
\frac{(|y|^2+1)}{|y|} \left(\pi - \arctan{\frac{1}{|y|}} \right)
\right] \label{p45}
\end{equation}
is positively defined. Again, we should eliminate the sign minus
in front of the square root in (\ref{p44}) since $k>0$. Thus, for
arbitrary $p$ and $k>0$ there are solutions $\Omega = i \gamma$
with negative $\gamma$, which describe damping transversal
nonharmonic perturbations in plasma.

\begin{figure}[t]
\includegraphics[width=8.0cm]{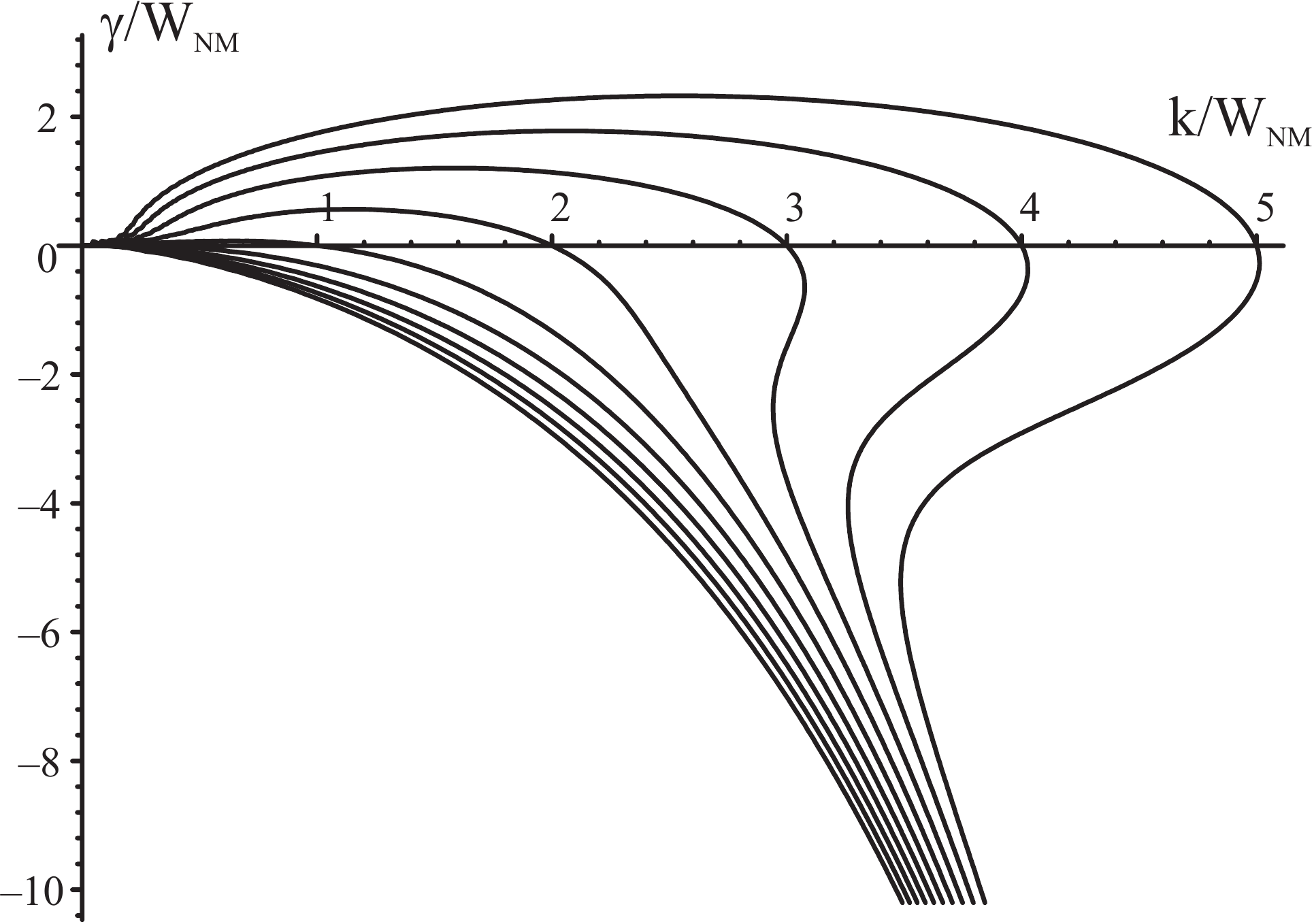}
\caption{Plots of the function $\gamma(k)$ for various values of
the parameter $p$. Ten curves  (indicated from the left to the
right) correspond to the values ${-}4,{-}3,\dots,4,5$ of the
reduced parameter $\frac{p}{W_{{\rm NM}}}$. Depending on the
chosen value of the parameter $k{=}k_0$, one can find one, two or
three solutions $\gamma(k_0)$.} \label{gtrans}
\end{figure}

Two asymptotic decompositions attract an attention
\begin{equation}
\gamma(k)=\frac{4p}{3\pi W^2_{{\rm NM}}}\,k^2+\dots, \qquad k\ll
W_{{\rm NM}}\,, \label{2p35}
\end{equation}
and
\begin{equation}
\gamma(k)=-\frac{2}{3\pi W^2_{{\rm NM}}} \,k^3+\dots, \qquad k\gg
W_{{\rm NM}}\,. \label{3p35}
\end{equation}
In Fig.~\ref{gtrans} we illustrate the dependence $\gamma(k)$ for
ten values of the parameter $p$; we put together the plots illustrating
the solutions discussed in two paragraphs: for $x{=}0$, $y>0$ and for $x{=}0$, $y<0$.

\begin{figure}[h]
\begin{center}
\begin{tabular}{ccc}
\includegraphics[height=5.2cm]{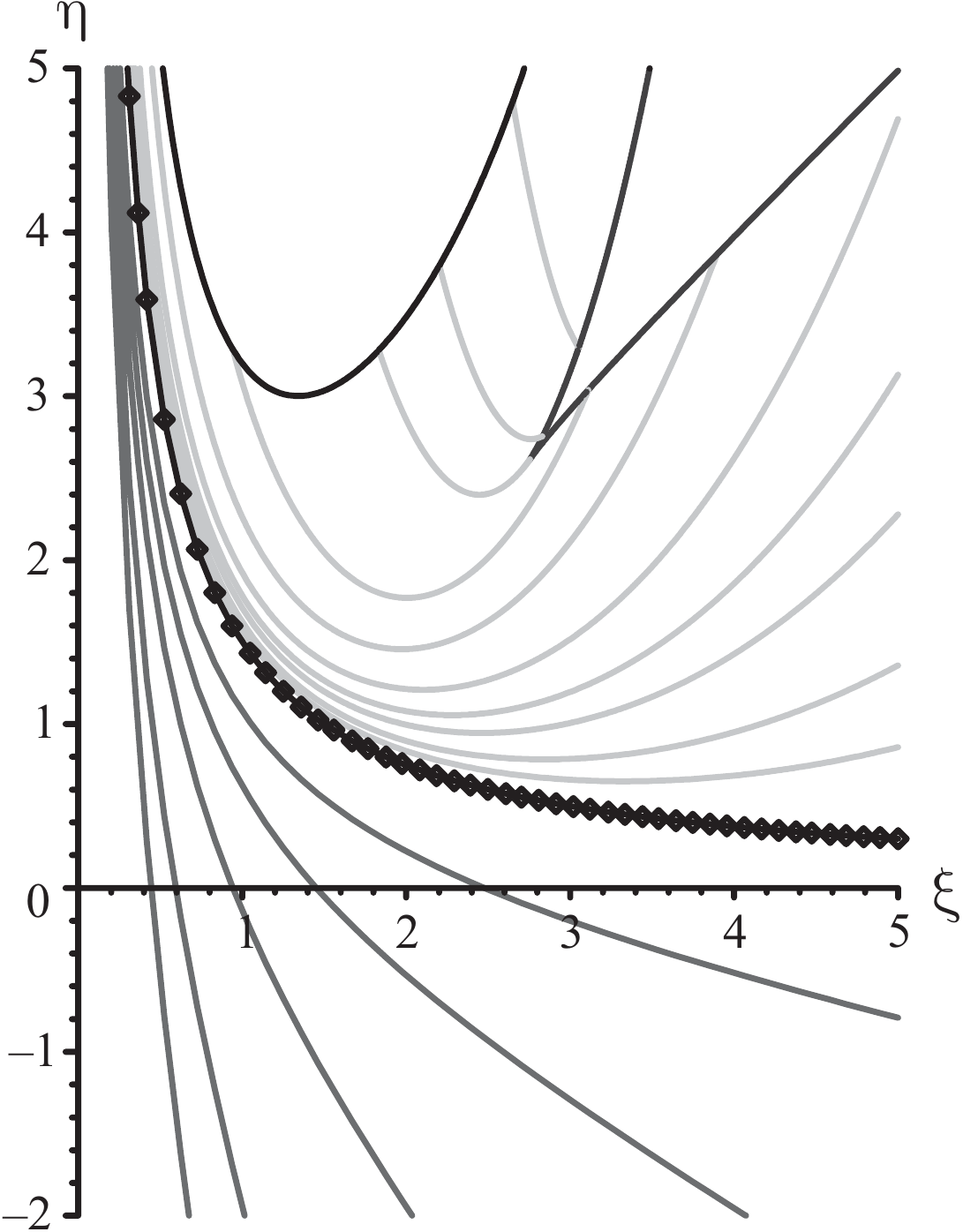}&\qquad&
\includegraphics[height=5.2cm]{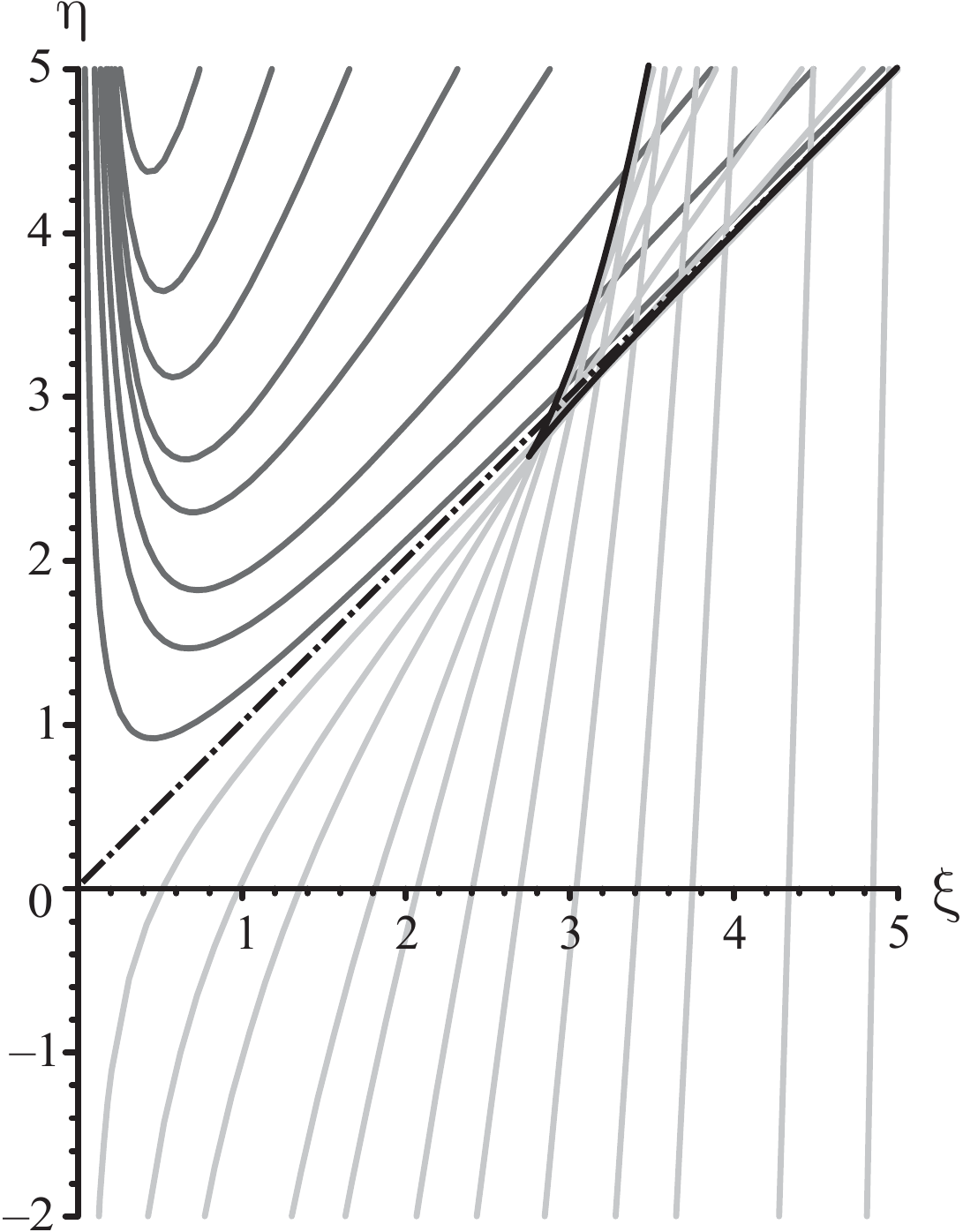}\\
a& & b\\
\end{tabular}
\end{center}\caption{
The curves $p=p(k)$ reconstructed using parametric equations
(\ref{7p1}) and (\ref{7p2}). On the left panel we illustrate
solutions of the dispersion equation for the transversal waves
with $\Real z>1$ and $\Im z =0$ (dark grey infinite lines), and
the solutions with $0<\Real z <1$ and $\Im z <0$ (light grey
finite lines). On the right panel we display the family of lines
related to the solutions with $\Real z {=}0$ and $\Im z >0$ (dark
grey), and to the solutions with $\Real z {=}0$ and $\Im z <0$
(light grey). The curves on the left panel visualize three
separatrices; the curves on the right panel add one new
separatrix; all these distinguished curves divide the plane $p0k$
into seven domains, as it is shown on Fig.~\ref{Fig1}.
}\label{plotI}
\end{figure}

\subsubsection{The case $x \neq 0$ and $y \neq 0$}

\noindent We indicate the solutions of the dispersion equations,
for which neither $\Real{z}$, nor $\Im{z}$ vanishes, as nontrivial
solutions, and study them numerically. For this purpose we
introduce the dimensionless real quantities $\xi =
\frac{k}{W_{{\rm NM}}}$,  $\eta =\frac{p}{W_{{\rm NM}}}$ and link
them by two parametric equations
\begin{gather}
\xi^2=\frac{3}{2} \frac{\Im\textsf{G}_{\bot}(z)}{\Im z^2}\,,
\label{7p1} \\
\xi \eta =\frac{3}{2}
\frac{\Im\left[\textsf{G}_{\bot}(z)(z^2-1)^{-1}\right]}{\Im
\left[(z^2-1)^{-1}\right]}\,. \label{7p2}
\end{gather}
We use the following scheme of analysis. We are interested,
finally, in the determination of the functions $\omega
{=}\omega(k,p)$ and $\gamma{=}\gamma(k,p)$; however, we start with
the numerical and qualitative analysis of the inverse functions
$k{=}k(x,y)$ and $p{=}p(x,y)$. In other words, first of all, we
find the domains on the plane of the parameters $\xi$ and $\eta$,
in which the corresponding solutions exist.

In order to illustrate this scheme let us consider auxiliary plots
on Fig.~\ref{plotI}.  On Fig.~\ref{plotI}a one can find the lines
of two types: first, infinite lines, which relate to solutions
with $\Real z>1$ and $\Im z =0$; second, finite lines, which
correspond to the solutions with $0<\Real z <1$ and $\Im z <0$.
The family of infinite lines visualizes the separatrix in the form
of hyperbola. The family of finite lines visualizes two curves on
which these lines start or finish. In Fig.~\ref{plotI}b one can
find infinite lines of two types. First, we see the family of
lines, for which $\Real z =0$ and $\Im z >0$; they visualize the
separatrix in the form of straight line. The second family of
lines relates to the solutions with $\Real z =0$ and $\Im z <0$;
they visualize an envelope line with extreme point.

\begin{figure}[t]
\includegraphics[width=7cm]{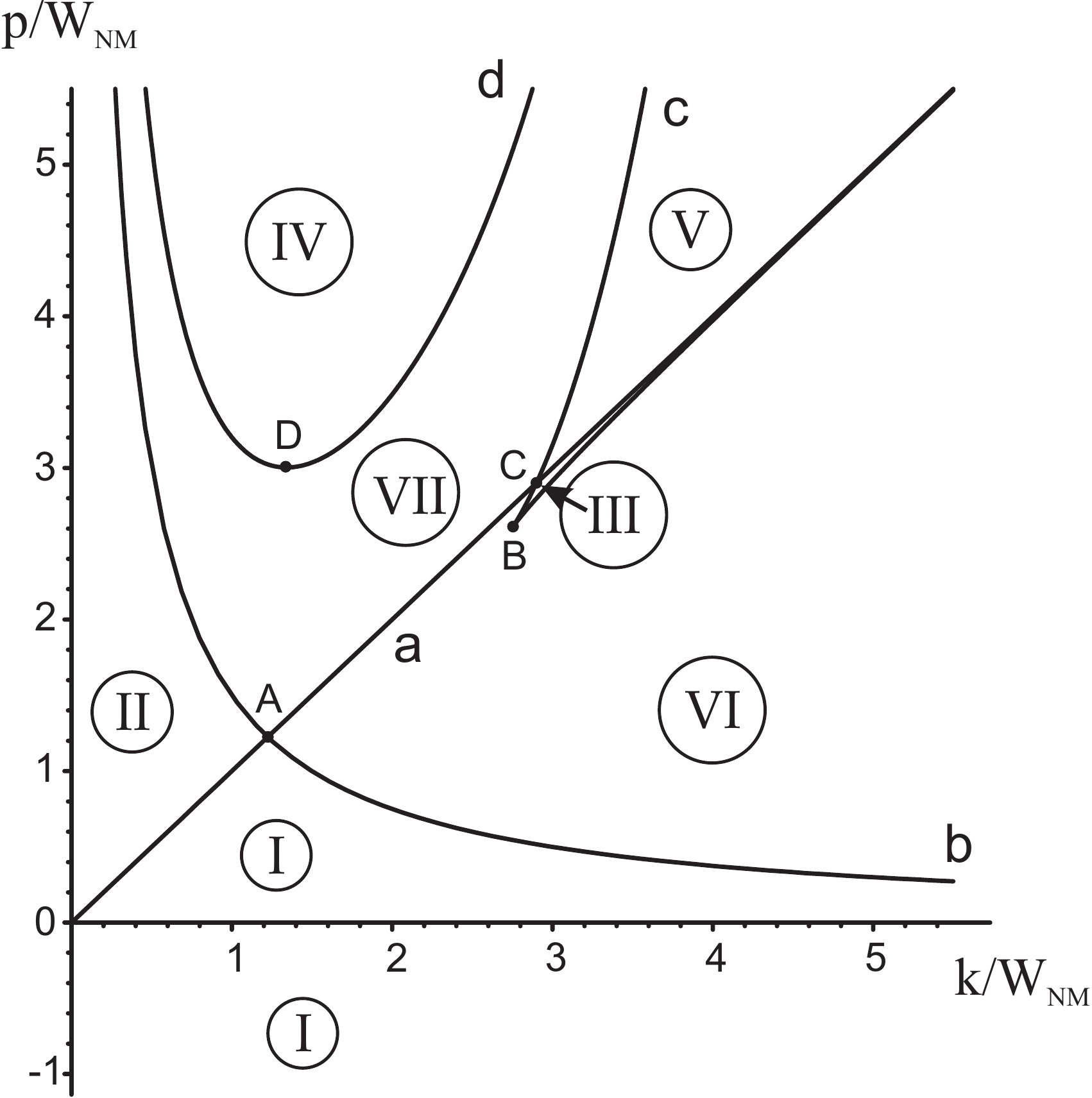}
\caption{Arrangement of domains on the plane $p0k$ with specific
types of solutions to the dispersion equation (\ref{transDR13}) at
$1+2K_1>0$. When the axion field is stationary ($\nu{=}0$), the
parameter $p{=}\mp\nu \frac{(1{+}2K_2)}{(1{+}2K_1)}$ is equal to
zero. In the domain I the solutions with $\gamma=0$, $\omega>k$
(running transversal waves), and the solutions with $\omega=0$,
$\gamma<0$ (damping nonharmonic perturbations) exist. In the
domain II there are running waves with $\gamma=0$, $\omega>k$, and
instable perturbations with $\omega=0$, $\gamma>0$. In the domains
III, IV and V there are nonharmonic perturbations only: in III
they are damping ($\gamma<0$), in IV they are instable
($\gamma>0$), in V both perturbations can be generated. In the
domains VI and VII damping waves with $\omega<k$ and $\gamma<0$
exist; in addition, there are damping perturbations in VI
($\gamma<0$) and instable ones in VII ($\gamma>0$).} \label{Fig1}
\end{figure}

We collected the results of numerical analysis on Fig.~\ref{Fig1}.
There are seven domains on the plane $\eta 0 \xi$, the
corresponding separatrices appeared as follows.

\noindent (c1) The line indicated as (a) is the bisector
$\xi{=}\eta$, and the appearance of this formula can be explained
as follows.  On the one hand, it relates to the straight-line
separatrix found by numerical calculations; on the other hand, it
corresponds to the special solution $z=0$, which appears if
$k=p>0$. In addition to the static solution $\omega{=}\gamma
{=}0$, in the points belonging to the bisector $p{=}k$ we obtain
the running waves with $\omega>k$, $\gamma{=}0$, when the wave
number $k$ satisfies the inequality $k<\sqrt{\frac{3}{2}}W_{{\rm
NM}}$.

\noindent (c2) The line (b) is described by the formula $\xi \eta
{=} \frac32$. On the one hand, is relates to the hyperbolic
separatrix found numerically; on the other hand it corresponds to
the critical case $k{=}k_{{\rm crit}}{=}\frac{3W^2_{{\rm
NM}}}{2p}$, found analytically. For the points along this
separatrix, the solutions $x=1$ exist, or equivalently,
$\omega{=}k$, $\gamma{=}0$; the solutions with $x^2>1$ do not
exist.

\noindent (c3) Keeping in mind that the nontrivial solutions to
the dispersion equations with $x \neq 0$ and $y \neq 0$ exist only
for $y<0$ and $|x|<1$, we associated the numerically found curves,
on which the finite lines start or finish, with two limiting lines
found analytically (for $x \geq 0$). The first curve of this type,
the curve (c), corresponds to the limit $x \to 0$ and is described
parametrically as
\begin{gather}
\xi^2 =\frac{3}{2} \lim_{x\to 0}\frac{\Im\textsf{G}_{\bot}(z)}{\Im
z^2} =
\frac{9}{4}\left(1+\frac{3|y|^2{+}1}{3|y|}\arccot(-|y|)\right),
\nonumber\\
\xi \eta =\frac{3}{2}\lim_{x\to
0}\frac{\Im\left[\textsf{G}_{\bot}(z)(z^2-1)^{-1}\right]}{\Im
\left[(z^2-1)^{-1}\right]}={}\nonumber\\
{}=\frac{3}{4}\left(|y|^2+3+\frac{(|y|^2+1)^2}{|y|}\arccot(-|y|)\right)\,.
\label{7p3}
\end{gather}
The second curve, the curve (d), relates to the limit $x \to 1$
and has the following parametric representation:
\begin{gather}
\xi^2=\frac{3}{2} \lim_{x\to
1_{-0}}\frac{\Im\textsf{G}_{\bot}(z)}{\Im z^2},\nonumber\\ \xi
\eta =\frac{3}{2}\lim_{x\to
1_{-0}}\frac{\Im\left[\textsf{G}_{\bot}(z)(z^2-1)^{-1}\right]}{\Im
\left[(z^2-1)^{-1}\right]}. \label{7p4}
\end{gather}
Clearly, in the points of the curve (c) the solutions to the
dispersion relations have the form $\omega=0$, $\gamma <0$; as for
the curve (d), one obtains that $\omega=k$, $\gamma<0$.

In the cross-points the solutions inherit the properties, which
characterize both separatrices.

\noindent (p1) The lines (a) and (b) cross in the point A with the
coordinates $\eta = \xi = \sqrt{\frac32} \simeq 1.225$; in this
point $\omega=\gamma=0$ or $\omega=k=\sqrt{\frac{3}{2}}W_{{\rm
NM}}$, $\gamma=0$.

\noindent (p2) The lines (a) and (c) cross in the point C with the
coordinates $\eta = \xi  \simeq 2.899$; in this points there are
two solutions: $\omega=\gamma=0$ and $\omega=0$, $\gamma=-2.32\,
W_{{\rm NM}}$.

\noindent (p3) The extreme point B has the coordinates $\xi \simeq
2.754$, $\eta \simeq 2.613$; for this point there exists a
solution with $\omega=0$, $\gamma=-1.31 W_{{\rm NM}}$.

\noindent (p4) Finally, the point D, which marks the minimum of
the curve (d) has the coordinates $\xi \simeq 1.342$, $\eta \simeq
3.003$; in this point the special solution $\omega=k$,
$\gamma=-0.303\,W_{{\rm NM}}$ exists.

Let us summarize the properties of solutions, which relate to the
domains I, II, \ldots, VII in Fig.~\ref{Fig1}, using the analysis
of the formulas presented above.
\begin{itemize}
\item The domain I is characterized by the inequalities $\eta<\xi$
and $\eta <\frac{3}{2\xi}$, or in other terms $p<k$ and $k <
k_{{\rm crit}} {=}\frac{3W^2_{{\rm NM}}}{2p}$; this domain
includes the region $p \leq 0$. In this domain, first, the
nontrivial solutions with $x \cdot y \neq 0$ do not exist; second,
there are running waves without damping/increasing ($\gamma
{=}0$), which are characterized by the phase velocity exceeding
the speed of light in vacuum ($\omega > k$); third, the damping
nonharmonic perturbations exist with $\omega {=}0$ and $\gamma
<0$. The line $\eta{=}0$, which belongs this domain, relates to
the model without axions ($\nu{=}0$ and thus $p{=}0$), and in this
sense our results recover the well-known ones.

\item The domain II ($\eta>\xi$ and $\eta <\frac{3}{2\xi}$)
accumulates solutions with $p>k$ and $k < k_{{\rm crit}}$; the
corresponding solutions differ from the ones in the domain I by
one detail only: the nonharmonic perturbations with $\omega {=}0$
are instable, since now $\gamma >0$ (see the discussion below the
formula (\ref{p35})).

\item In the domains III, IV and V only nonharmonic perturbations
exist: the nontrivial solutions with $x \neq 0$, $y \neq 0$ are
absent, the solutions with $x \neq 0$, $y {=} 0$ are not
admissible since $k>k_{{\rm crit}}$ here. The damping nonharmonic
solutions with $\omega {=}0$ and $\gamma<0$ are admissible in III
and V, the instable solutions with $\gamma>0$ exist in IV and V.

\item Non-trivial solutions with $x \neq 0$, $y \neq 0$ are
admissible in the domains VI and VII; here the damping waves exist
with $\omega<k$ and $\gamma<0$. In addition nonharmonic
perturbations are admissible, which are damping ($\gamma <0$) in
VI, and increasing ($\gamma >0$) in VII.
\end{itemize}
These results are presented shortly in Table~\ref{table1}.

Let us attract the attention to the number of symbols $+$ in the
corresponding boxes. For instance, there are three symbols $+$ in
the box for the domain III with negative $\gamma$, two symbols  in
the domain V, and one plus in the domains I and VI. This means
that for one value of the quantity $k$ we obtain one, two or three
values of the decrement of damping $\gamma(k)$. This feature can
be explained using Fig.~\ref{gtrans}: depending on the value of
the parameter $p$ the vertical straight line $k{=}{\rm const}$ can
cross the curve $\gamma{=}\gamma(k,p)$ one, two or three times,
respectively.
\begin{table}[h]
\begin{center}
\caption{$1+2K_1>0$}\label{table1} 
\begin{tabular}{||c|c|c|c|c||}
  \hline
  \,Domain  & $\Omega=\omega$ & \,$\Omega=-i|\gamma|$\, & $\Omega=i|\gamma|$ & $\Omega=\omega-i|\gamma|$ \\
     & ($\omega>k$) &  &  & ($\omega<k$) \\
  \hline\hline
  I & + & + & --- & --- \\
  \hline
  II & + & --- & + & --- \\
  \hline
  III & --- & +\,+\,+ & --- & ---  \\
  \hline
  IV & --- & --- & + & ---  \\
  \hline
  V & --- & +\,+ & + & --- \\
  \hline
  VI & --- & + & --- & + \\
  \hline
  VII & --- & --- & + & + \\
  \hline
\end{tabular}
\end{center}
\end{table}

\subsection{The third case: $1+2K_1<0$}

\begin{figure}[h]
\includegraphics[width=7cm]{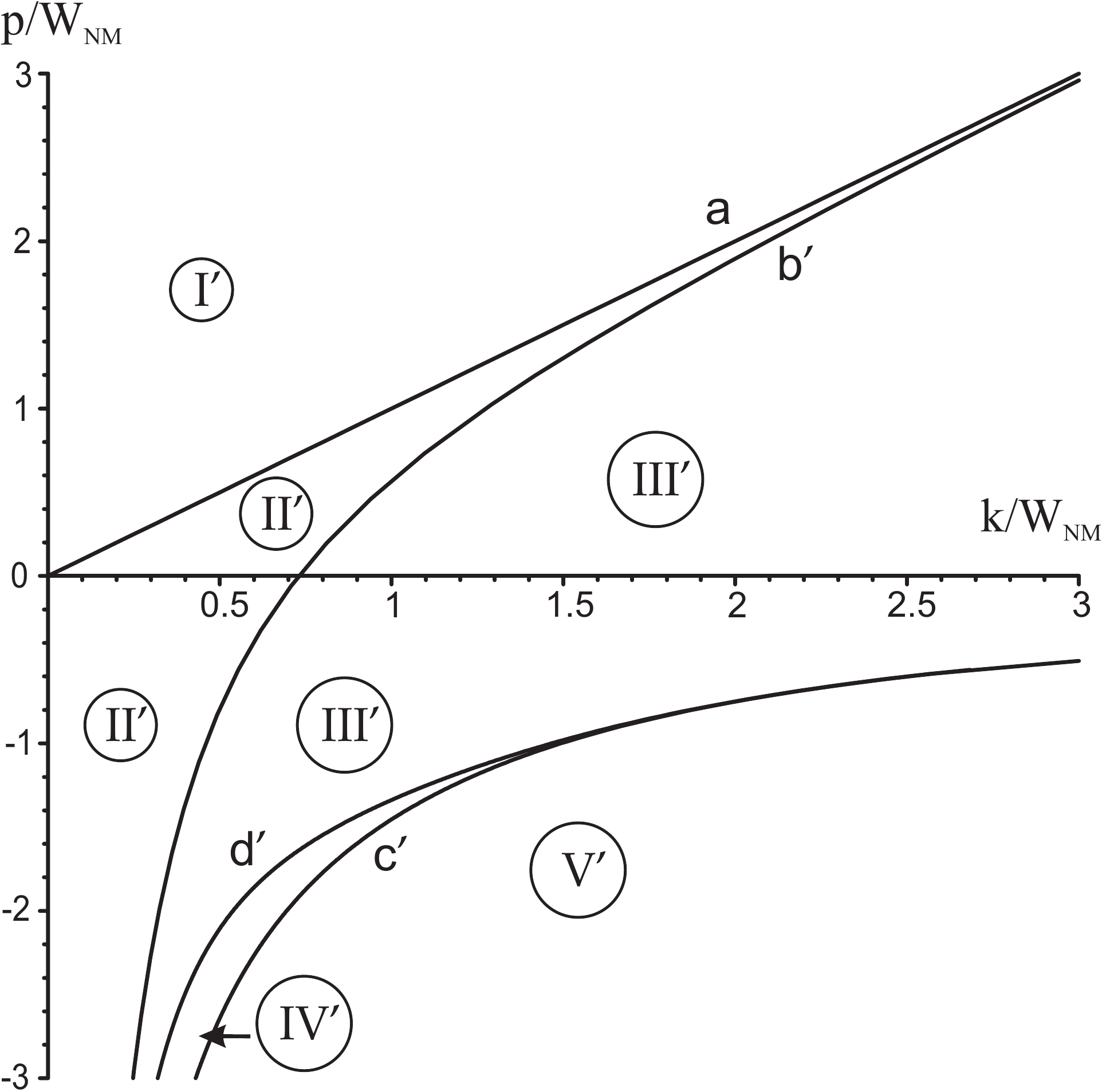}
\caption{ Arrangement of domains with specific types of solutions
to the dispersion equation (\ref{transDR13})  at $1+2K_1<0$. In
the domains I$^\prime$ and II$^\prime$ all the solutions belong to
the class of nonharmonic perturbations with $\omega=0$; in
I$^\prime$ these perturbations are damping ($\gamma<0$), while in
II$^\prime$ they are instable ($\gamma>0$). The domain
III$^\prime$ contains instable waves with $\omega<k$ and
$\gamma>0$; in the domain IV$^\prime$ there exist both damping and
instable waves with $\omega>k$. The domain V$^\prime$ contains the
solutions for running waves with $\gamma=0$, $\omega>k$. }
\label{Fig2}
\end{figure}

This case can be obtained if we make the formal replacement
$W_{{\rm NM}}^2 \to - W_{{\rm NM}}^2$ in all formulas obtained in
the previous subsection. In particular, the dispersion relation
for the transversal waves in plasma has now the form
\begin{equation}
\xi^2(z^2-1)+ \xi \eta +\frac{3}{2} \,\textsf{G}_{\bot}(z)=0 \,.
\label{transDR4}
\end{equation}
We do not discuss similar details of the corresponding analysis,
and present below the results only. Similarly to the case
$1{+}2K_1>0$, numerical modeling of the lines $k{=}k(p)$ displays
three separatrices on the plane $\eta 0 \xi$; they are shown in
Fig.~\ref{Fig2}. The line (a), again, is the bisector
$\xi{=}\eta$. The parametric representation of the line ($\rm
b^\prime$) is
\begin{gather}
\xi^2=\frac{9}{4}
\,\left[\frac{3y^2+1}{3y}\arccot(y)-1\right]\,,\nonumber\\
\xi \eta =\frac{3}{4}\,\left[\frac{(y^2+1)^2}{y}\arccot(y)-y^2-3\right]\,.
\label{env1}
\end{gather}
It is obtained using the conditions $x{=}0$ and $y>0$ (see
(\ref{7p3})). The line ($\rm c^\prime$) is described by the parametric equations
$$
\xi^2=\frac{9}{4}\,\left[\frac{3x^2-1}{6x}\ln \left(\frac{x+1}{x-1}\right)-1\right]\,,
$$
\begin{equation}
\xi \eta=\frac{3}{4}\,\left[x^2-3-\frac{(x^2-1)^2}{2x}\ln \left(\frac{x+1}{x-1}\right)\right]\,,
\label{env2}
\end{equation}
and is the envelope of the family of curves $p=p(x,y{=}0)$,
$k=k(x,y{=}0)$, when $x>1$. The last line (d$^\prime$) is obtained
using the conditions $x{=}1$ and $y>0$. Contrary to the previous
case, the separatrices have neither cross-points, nor extreme
points. Thus, in the case of negative constant $1{+}2K_1$, the
plane $\xi 0 \eta$ is divided into five domains. In the domain
I$^{\prime}$ the solutions to the dispersion equation have
vanishing real parts $\omega{=}0$ and negative imaginary parts
$\gamma{<}0$; the domain II$^{\prime}$ is characterized by
$\omega{=}0$ and $\gamma{>}0$. In the domain III$^{\prime}$ there
exist instable waves with $\omega <k$ and $\gamma>0$; the domain
IV$^\prime$ contains waves with $\omega >k$ of two types:
increasing ($\gamma>0$) and damping ($\gamma<0$). The domain
V$^{\prime}$ describes running waves  without damping/increasing,
$\omega>k$, $\gamma {=}0$. We summarize these features in
Table~\ref{table2}.

\begin{table}
\begin{center}
\caption{$1+2K_1<0$}\label{table2} 
\begin{tabular}{||c|c|c|c|c|c||}
  \hline
  \,Domain  & $\Omega=\omega$ & \,$\Omega{=}{-}i\gamma$\, & $\Omega=i\gamma$ & $\Omega=\omega{+}i\gamma$ & $\Omega=\omega{+}i\gamma$ \\
     & ($\omega>k$) & ($\gamma>0$) & ($\gamma>0$) & ($\gamma>0$) & ($\omega{>}k,\; \gamma{<}0$) \\
  \hline\hline
  I$^\prime$ & --- & + & --- & --- & --- \\
  \hline
  II$^\prime$ & --- & --- & + & --- & --- \\
  \hline
  III$^\prime$ & --- & --- & --- & + & --- \\
  \hline
  IV$^\prime$ & --- & --- & --- & --- & + \\
  \hline
  V$^\prime$ & + & --- & --- & --- & --- \\
  \hline
\end{tabular}
\end{center}
\end{table}

Concerning the principally new result, let us stress, that after
the replacement $W_{{\rm NM}}^2 \to {-} W_{{\rm NM}}^2$ in
(\ref{tr45}) we can see that indeed, in addition to $x{=}0$ and
$y{=}0$,  new solutions can exist with $x \neq 0$ and $y \neq 0$.
This means that, when $1{+}2K_1<0$, transversal waves in
axionically active plasma can be, first, damping waves with
$\omega>k$, $\gamma<0$, second, can be instable ($\gamma>0$) both
for $\omega>k$ and $\omega<k$.

\subsection{The intermediate case $1{+}2K_1=0$}

To complete our study, let us consider the case
$K_1{=}{-}\frac12$, for which the dispersion relation reduces to
the equation
\begin{equation}\label{transDR5}
k p_* = \frac32 W^2 \,\textsf{G}_{\bot}(z)\,,
\end{equation}
where $p_*= \mp \nu (1+2K_2)$ and $W$ is given by (\ref{an3}). In
analogy with the case $1+2K_1>0$, four separatrices appear on the
plane $\xi 0 \eta$ (here $\xi=\frac{k}{W}$, $\eta=\frac{p_*}{W}$).
The first separatrix is the hyperbola $\xi \eta =1$; it
corresponds to $\textsf{G}_{\bot}(\infty)=\frac23$ (minimal value
of the function $\textsf{G}_{\bot}(x)$). The second separatrix
$\xi \eta =\frac32$ corresponds to $\textsf{G}_{\bot}(1){=}1$
(maximal value of $\textsf{G}_{\bot}(x)$). The third separatrix
has the following representation:
\begin{equation}
\xi\eta=2.3904,
\end{equation}
where the number $2.3904$ is the
maximum of the function $\frac32\Real\{\textsf{G}_{\bot}(z)\}$ at the
condition that $\Im\{\textsf{G}_{\bot}(z)\}=0$.

\begin{figure}[h]
\includegraphics[width=7cm]{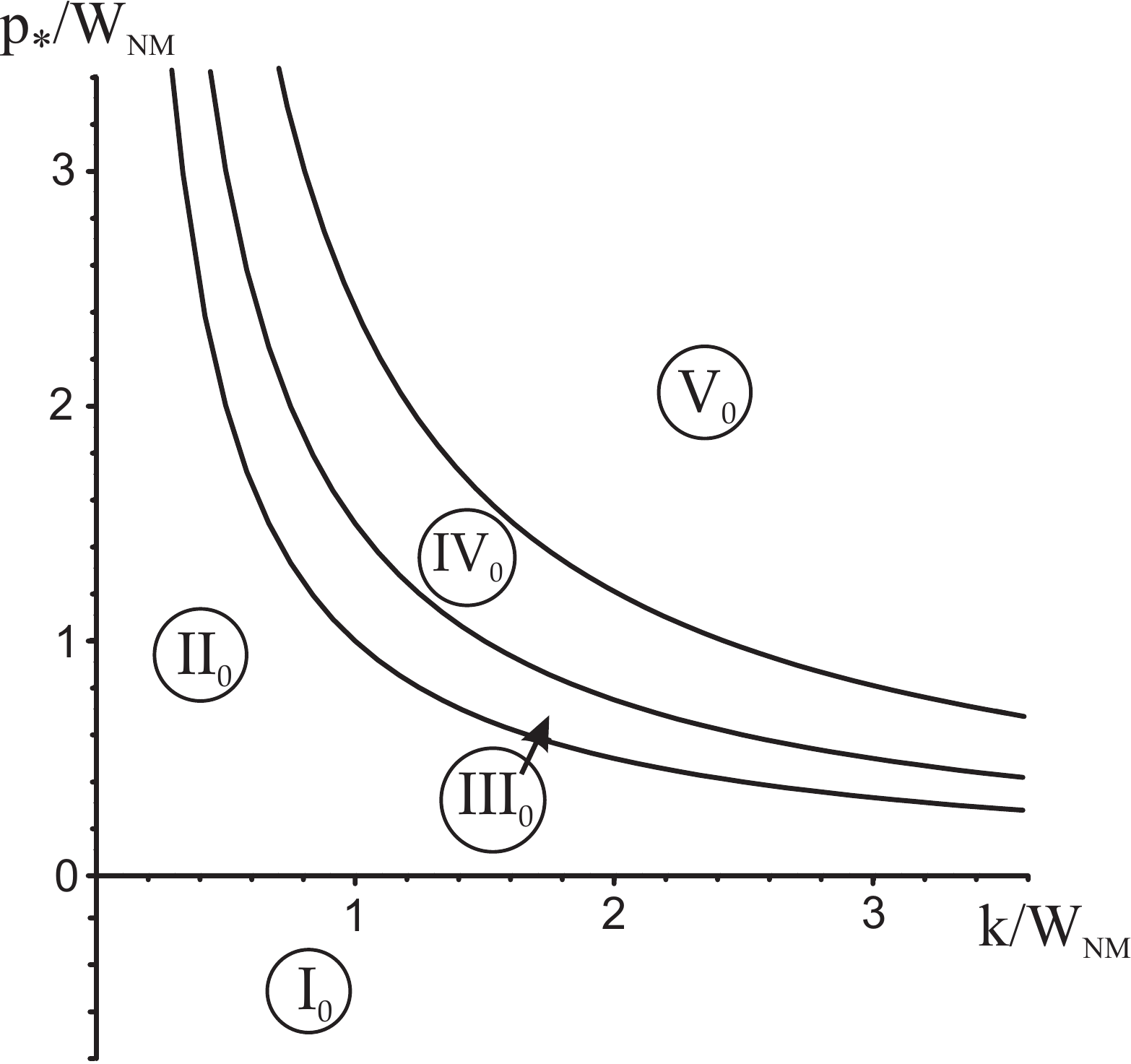}
\caption{ Arrangement of domains with specific types of solutions
to the dispersion equation (\ref{transDR13})  at $1+2K_1{=}0$.
Here $p_{*}=\mp\nu (1+2K_2)$ is the modified axionic guiding
parameter. In the domain I$_0$ only solutions with $\gamma<0$
exist; in the domain II$_0$ the solutions with $\omega{=}0$,
$\gamma>0$ are possible. In the domain III$_0$ there are running
waves with $\omega>k$. The domain IV$_0$ is characterized by the
damping waves with $\omega<k$ and $\gamma<0$. In the domain V$_0$
there are no solutions. }\label{Fig3}
\end{figure}

Finally, we visualize the straight line $p_*{=}0$ as an abscissa
on the plane $p_{*}0k$ and the last separatrice. Thus, we obtain
five domains on the plane $\xi 0 \eta$ (see Fig.~\ref{Fig3}). The
domain I$_0$ ($\eta<0$) is characterized by $\omega{=}0$ and
$\gamma<0$; in the domain II$_0$ ($\xi \eta <1$, $p_*>0$) the
solutions of the dispersion equations give $\omega{=}0$ and
$\gamma>0$. In the domain III$_0$ with $ 1<\xi \eta <\frac32$ we
obtain that $\gamma {=}0$ and $\omega>k$. The solutions with
$\omega<k$ and $\gamma <0$ can appear in the domain IV$_0$; in the
domain V$_0$ there are no solutions to the dispersion equation.

These results are summarized in Table~\ref{table3}.

\begin{table}
\begin{center}
\caption{$1+2K_1=0$}\label{table3} 
\begin{tabular}{||c|c|c|c|c||}
  \hline
  \,Domain  & $\Omega=\omega$ & \,$\Omega=-i\gamma$\, & $\Omega=i\gamma$  & $\Omega=\omega{+}i\gamma$ \\
     & ($\omega>k$) & ($\gamma>0$) & ($\gamma>0$) & ($\omega<k,\ \gamma<0$) \\
  \hline\hline
  I$_0$ & --- & + & --- & --- \\
  \hline
  II$_0$ & --- & --- & + & ---  \\
  \hline
  III$_0$ & + & --- & --- & ---  \\
  \hline
  IV$_0$ & --- & --- & --- & +  \\
  \hline
  V$_0$ & --- & --- & --- & ---  \\
  \hline
\end{tabular}
\end{center}
\end{table}

\section{Discussion}\label{sec6}

In the framework of nonminimal Einstein-Maxwell-Vlasov-axion model
we analyzed the dispersion relations for the perturbations in an
initially isotropic and homogeneous axionically active plasma,
which expands in the de Sitter-type cosmological background. In
this model we take into account, first, the nonminimal interaction
of the electromagnetic field with curvature, second, the
nonminimal coupling of the axion field to gravity, and
axion-photon coupling in the relativistic plasma. The specific
choice of the nonstationary background solution to the total
system of equations (de Sitter spacetime with constant curvature,
axion field linear in time, vanishing initial macroscopic
collective electromagnetic field in the electro-neutral
ultrarelativistic plasma) allowed us to obtain and study the
dispersion relations in the standard $(\Omega,\vec{k})$ form. We
consider these dispersion relations as nonminimal axionic
extension of the well-known dispersion relations obtained earlier
in the framework of the relativistic Maxwell-Vlasov plasma model.

The presence of the pseudoscalar (axion) field provides the plasma
to become a gyrotropic medium, which displays the phenomenon of
optical activity. When the plasma in the axionic environment is
initially spatially isotropic, its three-dimensional gyration
tensor, ${\cal G}_{\alpha \beta \gamma}$ is proportional to the
three-dimensional Levi-Civita tensor, $\epsilon_{\alpha \beta
\gamma}$, thus providing the optical activity to be of the natural
type \cite{LLVIII}. The proportionality coefficient is the
pseudoscalar quantity $p=\mp \nu \ \frac{(1{+}2K_2)}{(1{+}2K_1)}$.
The upper sign of this coefficient relates to the circularly
polarized wave with left-hand rotation; formally speaking, the
sign of $p$ depends on the sign of $\nu=\dot{\phi}$, and on the
sign of nonminimal coefficient $\frac{(1{+}2K_2)}{(1{+}2K_1)}$.
Since $\nu =\dot{\phi}$, this type of optical activity is induced
by the pseudoscalar (axion) field; if the axions form the dark
matter, the quantity $\nu$ is proportional to the square root of
the dark matter energy density (see, e.g., \cite{ax1,ax2} for
details). When $K_1 \neq 0$ and $K_2 \neq 0$ (see their
definitions in (\ref{eld7}), the gyration coefficient $p$ contains
the square of the Hubble function (i.e., the spacetime curvature
scalar in the case of the de Sitter model). Thus, we deal with
combined axionic-tidal gyration effect. The frequencies of
transversal electromagnetic waves are shown to depend not only on
the wavelength, but also on the gyration coefficient $p$ (see,
e.g., (\ref{appr1})), and this dependence has a critical
character. To be more precise, when $p{\neq}0$, the dispersion
equations admit some new branches of solutions
$\omega=\omega(k,p)$, $\gamma =\gamma(k,p)$ in addition to the
standard ones. If to consider the transversal electromagnetic wave
propagation in terms of left- and right-hand rotating components,
one can state, that one of the waves (say, with left-hand
rotation) can have arbitrary wavelength, while the second wave can
possess the wave number less than critical one (see, e.g.,
(\ref{p141})); in this sense we deal with some kind of mode
suppression caused by the axion-photon interactions.

In order to simplify the classification of the electromagnetic
modes in an axionically active plasma, we use the following
terminology: {\it damping wave}, when $\omega \neq 0$, $\gamma<0$;
{\it instable wave}, when $\omega \neq 0$, $\gamma>0$; {\it
running wave}, when $\omega \geq k$, $\gamma=0$; {\it damping
nonharmonic perturbation}, when $\omega = 0$, $\gamma<0$; {\it
instable nonharmonic perturbation}, when $\omega = 0$, $\gamma>0$.
The results of analytic, qualitative and numerical study are
presented on Figs.~\ref{Fig1}-\ref{Fig3} and in Tables
\ref{table1}-\ref{table3}. Based on these results, we can
illustrate the evolution of types of the transversal
electromagnetic perturbations depending on the value of the
gyration parameter $p$ for small, medium and large $k$. For this
purpose we draw, first, the horizontal straight line on
Fig.~\ref{Fig1}, and move it from the bottom to upwards.

\begin{itemize}
\item When $p\leq 0$, the horizontal straight line does not cross
any separatrices; in this case for arbitrary $k$ there exist
solutions of two types: first, running waves with $\omega>k$,
$\gamma =0$, or in other words $V_{{\rm ph}}>1$; second, damping
nonharmonic perturbations.

\item When $0<p<\sqrt{\frac32}W_{{\rm NM}}$ (below the cross-point
A), the horizontal straight line crosses the separatrices (a) and
(b), thus displaying three zones. In the zone of medium $k$
($p<k<\frac{3W^2_{{\rm NM}}}{2p}$) the running waves with $V_{{\rm
ph}}>1$ and damping nonharmonic perturbations happen to be
inherited from the zone $p\leq 0$; in the zone of small $k$
($k<p$) the running waves with $V_{{\rm ph}}>1$ are inherited, but
damping nonharmonic perturbations convert into the instable ones;
in the zone of large $k$ ($k>\frac{3W^2_{{\rm NM}}}{2p}$) the
running waves convert into damping waves, but damping nonharmonic
perturbations are inherited.

\item When $1.225<\frac{p}{W_{{\rm NM}}}<2.613$ (between
cross-point A and extreme point B), there are also three zones.
When $k<\frac{3W^2_{{\rm NM}}}{2p}$, there exist running waves and
instable nonharmonic perturbations; when $\frac{3W^2_{{\rm
NM}}}{2p}<k<p$, running waves convert into damping waves and
instable nonharmonic perturbations are inherited; when $k>p$,
damping waves and damping perturbations are inherited.

\item Next interval $2.613<\frac{p}{W_{{\rm NM}}}<2.899$ can be
describe similarly; new zone indicated as III appears, in which
damping waves convert into damping  nonharmonic perturbations.

\item When $2.899<\frac{p}{W_{{\rm NM}}}<3.003$, the following new
detail appears: the zone of medium $k$ breaks up into three
subzones, and in one of them damping waves appear instead of
running waves (in the domain VII).

\item When $p > 3.003 W_{{\rm NM}}$, the description is similar,
but the new subzone appears (see domain IV), in which waves do not
exist.

\end{itemize}
Similarly, we can illustrate the evolution of the modes for the
case $1+2K_1<0$ and  $1+2K_1{=}0$. Such analysis reveals two
interesting features. First, when the gyration parameter $p$ and
the nonminimal parameter $1{+}2K_1$ are fixed, we can find
explicitly the range for the wave parameter $k$, for which the
running waves, i.e., non-damping transversal electromagnetic
waves, can propagate in the Universe in the axionic dark matter
environment and bring us a true information about the Universe
structure and history. For instance, when  $1{+}2K_1>0$ according
to the Table~\ref{table1} it is possible in the domains I and II
only; thus, if the gyration parameter exceeds, say, the value
$p{=} \sqrt{\frac32} W_{{\rm NM}}$ (see the point A on
Fig.~\ref{Fig1}), the running waves can propagate only in the
narrow interval of small $k$ (the interval of long waves). When
$1{+}2K_1<0$, according to Table~\ref{table2} the running waves
can propagate in the domain V$^{\prime}$ only; thus, if the
parameter $p$ is positive, the running transversal waves are
suppressed for any $k$. Second, we can focus on the problem of
existence of transversal electromagnetic waves of a new  type,
i.e., transversal waves with the phase velocity less that the
speed of light in vacuum, $\omega<k$, which can interact in a
resonant manner with particles co-moving with them. It is
well-known that in the minimal theory ($K_1=0$) at the absence of
axion field $\nu=0$ the dispersion equations do not admit the
transversal waves of this type. Our consideration shows that the
damping waves with $V_{{\rm ph}}<1$ exist in the domains VI and
VII (see Fig.~\ref{Fig1} for the case $1+2K_1>0$), and the
instable waves with $V_{{\rm ph}}<1$ can be generated for the case
$1+2K_1<0$. Since the phase velocity of the transversal
electromagnetic wave is less than speed of light in vacuum, we can
make the following remark. There are plasma particles, which
co-move with this transversal electromagnetic wave, thus providing
a resonant interaction; in the case $1+2K_1>0$ this resonant
interaction leads to the Landau-type damping, since the wave
transfers the energy to the resonant particles; in the case
$1+2K_1<0$ this resonant interaction provides the Landau-type
instability, since now resonant particles transfer their energy to
the transversal plasma wave.

In the next paper we plan to study the dispersion relations for
the axionically active plasma nonminimally coupled to gravity in
the framework of cosmological Bianchi-I model with initial
magnetic field and electric field axionically induced.

\appendix

\acknowledgments The work was partially supported by the Russian
Foundation for Basic Research (Grants Nos. 11-02-01162 and
11-05-97518), by the Federal Targeted Program N14.T37.21.0668 and
the State Assignment N5.2971.2011.

\end{document}